\documentclass[final]{eps}

\usepackage{graphicx}
\usepackage{times}
\usepackage{amsmath,amssymb}

\volumenumber{00}
\publishyear{2013}
\frompage{\pageref{firstpage}}
\topage{\pageref{finalpage}}

\shorttitle{H. Hirashita and H. Kobayashi:
Shattering in the Interstellar Medium.}

\title{Evolution of dust grain size distribution by
shattering in the interstellar medium:
robustness and uncertainty}
\author{Hiroyuki Hirashita$^1$ and Hiroshi Kobayashi$^2$}
\affiliation{$^1$Institute of Astronomy and Astrophysics,
Academia Sinica, P.O. Box 23-141, Taipei 10617, Taiwan\\
             $^2$Department of Physics, Nagoya University, Nagoya, Aichi 464-8602, Japan}
\received{November 17, 2012}\revised{February 28, 2013}\accepted{March 12, 2013}
\abstract{
Shattering of dust grains in the interstellar medium is
a viable mechanism of small grain production in galaxies.
We examine the robustness or uncertainty in the theoretical
predictions of shattering. We identify
$P_1$ (the critical pressure above which
{the deformation destroys the original lattice structures})
as the most important quantity in determining the timescale
of small grain production, and confirm that the same
$P_1/t$ ($t$ is the duration of shattering) gives
the same grain size distribution [$n(a)$, where $a$ is
the grain radius] after shattering within a
factor of 3. The uncertainty in the fraction
of shocked material that is eventually ejected as fragments
causes uncertainties in $n(a)$ by a factor of 1.3 and 1.6
for silicate and carbonaceous dust, respectively.
The size distribution of shattered fragments have minor
effects as long as $\alpha_\mathrm{f}\lesssim 3.5$
(the size distribution of shattered fragments
$\propto a^{-\alpha_\mathrm{f}}$), since
the slope of grain size distribution $n(a)$ continuously
change by shattering and becomes consistent with
$n(a)\propto a^{-3.5}$. The grain velocities as a function of
grain radius can have an imprint
in the grain size distribution especially for
carbonaceous dust. We also show that
the formulation of shattering can be simplified without
losing sufficient precision.
}

\begin{document}
\label{firstpage}
\maketitle
\copyrighttext{}

\section{Introduction}

The evolution of dust in galaxies is important to
the understanding of galaxy evolution, since dust
{grains govern} some fundamental physical
processes in the interstellar medium (ISM). First, they
dominate the absorption and scattering of the stellar
light, affecting the radiative transfer in the ISM.
Second, the dust surface is the main site for the
formation of molecular hydrogen.
The former process is governed by the
extinction curve (absorption and scattering
coefficient as a function of wavelength;
Hoyle and Wickramasinghe, 1969;
Draine, 2003) and the latter
by the total surface area of dust grains
(Yamasawa \textit{et al}., 2011). Since the
extinction curve and the total grain surface area
both depend strongly on the grain size distribution,
clarifying the regulating mechanism of grain size distribution
is of particular importance in understanding those important
roles of dust.

Mathis \textit{et al}.\ (1977, hereafter MRN) show that
a mixture of silicate and graphite with a grain size
distribution [number density of grains per grain radius,
denoted as $n(a)$ in this paper]
proportional to $a^{-3.5}$
{(a grain size distribution with a power index of
$-3.5$ is called MRN grain
size distribution)}, where $a$ is the grain radius
($a\sim 0.001$--0.25 $\mu$m), reproduces the
Milky Way extinction curve. Pei (1992) shows that the
extinction curves in the Magellanic Clouds are also
explained by the MRN
grain size distribution with different
abundance ratios between silicate and graphite.
Kim \textit{et al.}\ (1994) and
Weingartner and Draine (2001) have applied a more
detailed fit to the Milky Way extinction curve in order
to obtain the grain size distribution. Although their
grain size distributions deviate from the MRN size
distribution, the overall trend from small to large
grain sizes roughly follows a power law
with an index near to $-3.5$. In any models that
fit the Milky Way extinction curve,
the existence of a large number of small
($a\lesssim 0.01~\mu$m) grains is required.
The existence of such small grains is further supported
by the mid-infrared excess of the spectral energy
distribution in the Milky Way
(e.g., D\'{e}sert \textit{et al}., 1990;
Draine and Li, 2001).

{
Dust grains formed by stellar sources, mainly
supernovae (SNe) and asymptotic giant branch (AGB)
stars, are not sufficient to explain the total dust mass
in the Milky Way ISM, because the timescale of dust
enrichment by these sources is much longer than
that of dust destruction by SN shocks
(e.g., McKee, 1989; Draine, 1995). Moreover,
both SNe and AGB stars are suggested to supply large
($a\gtrsim 0.1~\mu$m)
grains into the ISM and cannot be the dominant source of
small grains (Nozawa, \textit{et al.}, 2007;
H\"{o}fner, 2008; Mattsson and H\"{o}fner, 2011;
Norris \textit{et al.}, 2012). Thus, some interstellar
processes (or more precisely, non-stellar processes)
are necessary to explain the large
abundance of small grains.}

There are some possible processes that efficiently
modify the grain size distribution.
Hellyer (1970) shows that the collisional
fragmentation of dust grains finally
leads to a power-law grain size distribution
similar to the MRN size distribution (see also
Bishop and Searle, 1983;
Tanaka \textit{et al}., 1996).
Hirashita and Yan (2009) show that such a
fragmentation and disruption process (or shattering)
can be driven efficiently by turbulence in the
diffuse ISM. Dust grains are also processed by other
mechanisms. Various authors show that the
increase of the total dust mass in the Milky Way ISM
is mainly governed
by grain growth through the accretion of
gas phase metals onto the grains (we call
elements composing dust grains ``metals'')
(e.g., Dwek, 1998; Zhukovska \textit{et al.},
2008; Inoue, 2011; Asano \textit{et al.}, 2012).
Grain growth mainly works in the dense ISM, such
as molecular clouds. In
the dense ISM, coagulation
also occurs, making the grain sizes larger
(e.g., Hirashita and Yan, 2009;
Ormel \textit{et al}., 2009). In the diffuse
ISM phase,
interstellar shocks associated with supernova (SN)
remnants destroy dust grains,
especially small ones, by sputtering (e.g., McKee, 1989;
Nozawa \textit{et al}., 2006).
Shattering also occurs in SN shocks
(Jones \textit{et al.}, 1996).
All these processes above other than shattering
do not contribute efficiently to the increase of 
small ($a\lesssim 0.01~\mu$m) grain abundance.
Indeed, Asano \textit{et al}.\ (2013) show that, unless we
consider shattering, the grain size distribution is
biased to large ($a\sim 0.1~\mu$m) sizes.
Therefore, shattering is important in reprocessing
large grains into small grains.

{
Based on the formulation developed by
Hirashita and Yan (2009),
Hirashita \textit{et al.}\ (2010, hereafter H10) consider
small grain production by shattering in the context
of galaxy evolution. H10 assume that Type II SNe
(SNe II) are the source of the first dust grains,
which are biased to large sizes ($\gtrsim 0.1~\mu$m)
because small grains are destroyed in the shocked
region in the SNe II (Nozawa \textit{et al.}, 2007,
hereafter N07). Starting with the size distribution of SNe II dust
grains in N07, H10 solve the shattering equation
to calculate the evolution of grain size distribution.
They show that the velocity dispersions acquired by
the dust grains in a warm ionized medium (WIM) is
large enough for shattering to produce
a large abundance of small grains on
a short ($\lesssim 10$ Myr) timescale. As mentioned
above, even if we consider other sources of dust
such as the dust formation in the wind of AGB stars
and the accretion of metals onto grains
in the interstellar clouds, shattering is generally
necessary to produce small grains (Asano \textit{et al}.\ 2013).}

Considering that shattering is a unique mechanism that
produces small grains efficiently in the ISM, it is
important to clarify how robust or uncertain the
theoretical calculations of shattering are. Since there
are some basic physical parameters regulating
shattering, it is
crucial to clarify how they affect the grain size
distribution or to which parameter shattering is
sensitive. Although any models should commonly
contain those basic parameters, there could be some
uncertainties in
the formulation itself, or there could be unnecessary
complexity which cannot be constrained from
observations anyway. Thus, in this paper, we examine
the robustness of
shattering calculations for the interstellar dust
by changing some major parameters
in a model and comparing the
results between two different frameworks.

This paper is organized as follows.
In Section \ref{sec:model}, we explain the shattering
models and pick out some major parameters.
In Section \ref{sec:result}, we examine the variation
of the grain size distribution {due to changes
of} the major parameters.
In Section \ref{sec:discussion},
we discuss our results and implications for the
evolution of grain size distribution.
In Section \ref{sec:conclusion},
we give our conclusions.

\section{Overview of the models for shattering}
\label{sec:model}

{
As mentioned in Introduction, we use the framework
in H10 to test the robustness of a
shattering model. H10 consider the evolution
of grain size distribution through shattering
by solving a shattering equation, which treats
the grain--grain
collision rate and the redistribution of shattered
fragments in the grain size distribution.
They
assume that grains collide with each other
under the velocity dispersions induced by
dynamical coupling with interstellar turbulence.
The production of fragments in grain--grain
collisions is the key process that we focus on
below.
}

The dust grains are assumed to be
spherical with radius $a$.
Thus, the grain mass $m$ is related to the
grain radius by
\begin{eqnarray}
m=\frac{4}{3}\pi a^3\rho_\mathrm{gr},
\end{eqnarray}
where $\rho_\mathrm{gr}$ is the material density of
the grain. {The grain size distribution, $n(a)$,
is defined so that $n(a)\, da$ is the number
density of grains with radii between $a$ and $a+da$.}

\subsection{Initial grain size distribution}
\label{subsec:initial}

The size distribution of grains ejected from SNe II into
the ISM is adopted from N07. This size distribution
is used as the initial condition for the calculation of
shattering. N07 treated dust nucleation and
growth in a SN II, taking
into account the dust destruction by kinetic and thermal
sputtering in the shocked region. Thus, the grain size
distribution calculated by N07 is regarded as that
ejected from SNe II to the ISM (see also
Bianchi and Schneider, 2007). Following H10,
we adopt 20 M$_\odot$ as a representative progenitor mass,
and the unmixed case in which
the original onion-like structure of elements is
preserved, since the extinction features of carbon and
silicon, which are major grain components in the
unmixed case, are consistent with observations
(Hirashita \textit{et al}., 2005; Kawara \textit{et al}., 2010).
The formed grain species
are C, Si, SiO$_2$, Fe, FeS, Al$_2$O$_3$, MgO,
MgSiO$_3$, and Mg$_2$SiO$_4$. According to N07,
small grains with
$a\lesssim 0.02~\mu$m are trapped in the
shocked region and are efficiently destroyed by
thermal sputtering if the
ambient hydrogen number density,
$n_\mathrm{H}$, is larger than 0.1 cm$^{-3}$.
{In this paper, we adopt
$n_\mathrm{H}=1$ cm$^{-3}$
(Section \ref{subsec:shatter}). As shown in N07
(see also H10), most of the grains
have radii 0.1--1 $\mu$m. The initial grain
size distributions are also shown later in
Fig.\ \ref{fig:fM}.}

\begin{table*}[t]
\renewcommand{\arraystretch}{1.2}
\vspace{-.3cm}
\caption{Summary of {experimental} properties.}
\vspace{-.1cm}
\begin{center}
\begin{tabular}{@{}lccccc@{}}\hline
Species & $\rho_\mathrm{gr}$ & $c_0$
& $s$ & $v_\mathrm{shat}$ & $P_1\,^\mathrm{a}$ \\
& (g cm$^{-3}$) & (km s$^{-1}$) & & (km s$^{-1}$) & (dyn cm$^{-2}$)
 \\ \hline
Silicate & 3.3 & 5   & 1.2 & 2.7 & $3\times 10^{11}$\\
Carbonaceous dust & 2.2 & 1.8 & 1.9 & 1.2 & $4\times 10^{10}$\\
\hline
\multicolumn{6}{l}{$^\mathrm{a}$ $P_1$ is varied in Models D and E in
Table \ref{tab:model}.}\\
\multicolumn{6}{l}{{Note: All the quantities are determined
experimentally. See Jones \textit{et al.}\ (1996) and references
therein.}}\\
\end{tabular}
\end{center}
\label{tab:species}
\end{table*}

{
The normalization of the grain size distribution
is determined as follows. We consider a solar
metallicity environment as a representative
case. (For other metallicities, the timescale
of shattering just scales with the inverse of
metallicity.) We parameterize
the dust abundance by the oxygen abundance.
The oxygen mass produced by a SN II of
20~M$_\odot$ progenitor is
$m_\mathrm{O}=1.58$ M$_\odot$ according to
Umeda and Nomoto
(2002), whose data were adopted by N07.
The solar oxygen abundance is
assumed to be
$Z_{\mathrm{O},\odot} =5.6\times 10^{-3}$ by mass ratio
(Lodders, 2003).
Thus, for the solar oxygen abundance, the corresponding
dust-to-gas ratio is given by
$\mathcal{D}_0=Z_{\mathrm{O},\odot} m_\mathrm{d}/m_\mathrm{O}$,
where $m_\mathrm{d}$ is the dust mass ejected
from a single SN II (0.14 M$_\odot$; N07; H10).
Here we assume that both oxygen and dust are supplied
from SNe II. The grain size distribution is normalized
so that the
total grain mass density integrated for all the
size range, $\rho_\mathrm{dust}$, is equal to
$1.4n_\mathrm{H} m_\mathrm{H}\mathcal{D}_0$,
where $m_\mathrm{H}$ is
the mass of hydrogen atom, and the factor 1.4 is the
correction for the species other than hydrogen.
}

{
As mentioned in Introduction, additional contribution
from AGB stars does not change the following results
significantly as long as
AGB stars also supply large ($a\gtrsim 0.1~\mu$m)
grains (see also Asano \textit{et al.}, 2013).
In other words, the use of N07's results
for the initial condition is aimed at investigating
the important role of shattering in
the production of small grains from large grains.}

\subsection{Shattering}\label{subsec:shatter}

{
We calculate the shattering processes in
the same way as in H10 by solving the shattering
equation. The collision frequency between grains
with various sizes is determined by
grain-size-dependent velocity dispersions.
The shattering equation used in H10 is
based on Hirashita and Yan (2009) (originally taken
from Jones \textit{et al.}, 1994, 1996).
Shattering is assumed to take place if the relative
velocity between a pair of grains is larger than
the shattering threshold velocities, 
$v_\mathrm{shat}$ (2.7 and 1.2 km s$^{-1}$
for silicate and graphite, respectively; Jones
\textit{et al.}, 1996).}
The results are not sensitive to $v_\mathrm{shat}$
as long as the grain velocities driven by turbulence
is much larger than the threshold, but are rather
sensitive to the hardness of the grain materials as
shown later.

{We consider a range
$a_\mathrm{min}=3\times 10^{-8}$~cm (3 \AA) and
$a_\mathrm{max}=3\times 10^{-4}$~cm (3 $\mu$m)
for the grain radii.}
The grains passing through the boundary of the
smallest radius are removed from the calculation.
{
Most grain models adopt a minimum grain radius
of a few $\times 10^{-8}$ cm
(Weingartner \& Draine, 2001;
Guillet \textit{et al}., 2009) although applying
bulk material properties to such small grains
could cause a large uncertainty. However,
even if we adopt
$a_\mathrm{min}=10^{-7}$ cm, the results does not
change, except that the grain size distribution is
truncated at $10^{-7}$ cm. This is because shattering
occurs in a top-down manner in the grain sizes and
the production of grains with
$a\sim\mbox{a few}\times 10^{-8}$ cm by
shattering is never enough for these small grains
to have a large contribution to the total shattering
rate.
}

Although nine grain species are predicted
to form (Section \ref{subsec:initial}), the material
properties needed for the calculation of shattering are
not necessarily available for all the species. Thus, we
divide the grains into two groups: one is carbonaceous
dust and the other is all the other species of dust
(called ``silicate''),
and apply the relevant material quantities of graphite
and silicate, respectively. The material properties
of silicate and graphite are taken from
Jones \textit{et al}.\ (1996) and summarized in
Table \ref{tab:species}.

Since the shattering equation
is general enough, the major uncertainties can be
produced by the treatment of shattering fragments.
Thus, we explain how to treat shattering fragments
below.

We consider {two colliding grains} whose masses are
$m_1$ and $m_2$ (the former grain is called
target grain),
and estimate the total fragment mass in $m_1$
as a result of this collision.
{Note that we consider
a collision between $m_2$ and $m_1$ again in the
calculation and consider the fragments in $m_2$
(that is, we consider the same collision twice to
treat the fragments of each colliding grain).}
The total fragment mass is determined by the
mass shocked to the critical pressure
in the target (Jones \textit{et al}., 1996;
Hirashita and Yan, 2009):
\begin{eqnarray}
\frac{M_\mathrm{ej}}{m_2}=f_M\frac{M}{m_2}=
\frac{f_M(1+2\mathcal{R})}{2(1+\mathcal{R})^{9/16}}
\frac{1}{\sigma_r^{1/9}}\left(
\frac{\mathcal{M}_r^2}{\sigma_1\mathcal{M}_1^2}\right)^{8/9},
\label{eq:frag}
\end{eqnarray}
where $M_\mathrm{ej}$ is the total mass of fragments
ejected from $m_1$, $M$ is the mass shocked to the critical
pressure ($P_1$), above which the solid becomes
plastic {(that is, the deformation destroys the
original lattice structures)}, $f_M$ is the fraction of shocked mass that
is eventually ejected as fragments (the rest
remains in the grain),
$\mathcal{R}=1$ in the collision between
the same species (we only consider collisions between
the same species for simplicity),
$\mathcal{M}_r\equiv v/c_0$ ($c_0$ is the sound speed
of the grain material), $\mathcal{M}_1$
is the Mach number corresponding to the critical
pressure $P_1$:
\begin{eqnarray}
\mathcal{M}_1=\frac{2\phi_1}{1+(1+4s\phi_1)^{1/2}},
\end{eqnarray}
where $\phi_1\equiv P_1/(\rho_\mathrm{gr}c_0^2)$
and $s$ is a dimensionless material constant
that determines the relation between the shocked velocity
and the velocity of the shocked matter.
{For convenience, we define function $\sigma$ as}
\begin{eqnarray}
\sigma (\mathcal{M})\equiv
\frac{0.30(s+\mathcal{M}^{-1}-0.11)^{1.3}}{s+\mathcal{M}^{-1}-1},
\label{eq:sigma}
\end{eqnarray}
{and evaluate $\sigma_1=\sigma (\mathcal{M}_1)$ and
$\sigma_r=\sigma (\mathcal{M}_r/(1+\mathcal{R}))$ in
Eq.\ (\ref{eq:frag})}. If $M$ is larger than half of
the grain mass,
we assume that the whole grain is fragmented; i.e.,
$M_\mathrm{ej}=m_1$. This case is called catastrophic
disruption.
{Otherwise (i.e., for $M<m_1/2$), only a
fraction of $m_1$ ($f_MM$) is ejected as fragments.
This case is called cratering.
The relation between the total volume ($V$)
of shattered fragments and the critical pressure is
roughly given by the total energy $P_1V=\mbox{constant}$.
Indeed, the above equations tell us that $M\propto V$ is}
inversely proportional to $P_1$, if $4s\phi_1 >1$, which is
true for most of the cases considered in this paper
($\mathcal{M}_1^2\sim \phi_1$, so
$M_\mathrm{ej}\propto \phi_1^{-8/9}\propto P_1^{-8/9}$).
{
Since the shattering efficiency is proportional to
$M_\mathrm{ej}$, the timescale of shattering is
regulated by $P_1$.}
Considering that the material properties of astronomical grains
are uncertain, we vary $P_1$ (Section \ref{subsubsec:Pcr}).

The total fragment mass $M_\mathrm{ej}$ is
distributed with a grain size distribution
$\propto a^{-\alpha_\mathrm{f}}$.
Jones \textit{et al}.\ (1996) argue that
$\alpha_\mathrm{f}=3.0$--3.4, based on an analysis
of {the flow associated with the formation
of a crater} (see also Takagi \textit{et al}., 1984;
Nakamura and Fujiwara, 1991;
Nakamura \textit{et al}., 1994;
Takasawa \textit{et al}., 2011
for experimental results).
Unless the catastrophic disruption occurs,
we assume that the mass $m_1-M_\mathrm{ej}$ remains
as a single dust grain and is distributed in an
appropriate bin in the numerical calculation.
To avoid unnecessary complexity
caused by the choice of the upper and lower radii
of the fragments, we adopt simpler forms for the
smallest and largest fragments than H11, following
Guillet \textit{et al}.\ (2009):
\begin{eqnarray}
a_{f\mathrm{max}}=(0.0204f)^{1/3}a_1\, ,\label{eq:afmax}
\end{eqnarray}
where $a_1$ is the radius of the grain $m_1$
(i.e., the target grain),
$f\equiv M_\mathrm{ej}/m_1$
(shattered fraction of the target grain).
The minimum radius is assumed to be
$a_{f,\mathrm{min}}=a_\mathrm{min}=3\times 10^{-8}$~cm.
If
$a_{f,\mathrm{max}}<a_\mathrm{min}$, we remove
the fragments from the calculation.

{
The relative velocity in the collision is
estimated based on the grain velocity dispersion
as a function of grain radius $a$. We used the calculation
by Yan \textit{et al.}\ (2004), who consider the grain
acceleration by hydrodrag and gyroresonance
in magnetohydrodynamic turbulence,
and calculate the grain velocities achieved in various
phases of ISM.}
Among the ISM phases, we focus on the
warm ionized medium (WIM) to
investigate the possibility of efficient shattering in
actively star-forming environments.
We adopt $n_\mathrm{H} =1$ cm$^{-3}$
for the hydrogen number density
of the WIM. The
resulting grain size distribution is not sensitive
to $n_\mathrm{H}$ (H10). Indeed, if $n_\mathrm{H}$ is large,
the grain--grain collision rate (i.e., the shattering
efficiency) rises, while
dust grains are more destroyed in
SNe (i.e., the dust-to-gas ratio is lower).
Because of these compensating effects,
the resulting grain size distribution is not sensitive
to the gas density.
We adopt gas temperature
$T=8000$ K, electron number density
$n_\mathrm{e}=n_\mathrm{H}$, Alfv\'{e}n speed
$V_\mathrm{A}=20$ km s$^{-1}$
and injection scale of
the turbulence $L=100$ pc.
For grains with
$a\gtrsim 0.1~\mu$m, where most of
the grain mass is contained in our cases, the grain
velocity is
governed by gyroresonance. Since there is still
an uncertainty in the typical grain radius ($a_\mathrm{c}$)
above which gyroresonance effectively works, we also
address the variation of results by $a_\mathrm{c}$
(Section \ref{subsubsec:vel}).

{
The grain velocities given above are
velocity dispersions. In order to estimate the
relative velocity ($v_{12}$) between the grains with $m_1$
and $m_2$, whose velocity dispersions are $v_1$ and $v_2$,
respectively, and we average the size distributions
of fragments for four cases
$v_{12}=v_1+v_2$, $|v_1-v_2|$, $v_1$, and $v_2$
to take the directional variety into account.
See H10 for more details.
}

\subsection{Duration of shattering}

{
Since we consider shattering in the WIM, it is
reasonable to assume that the shattering
duration (denoted as $t$) is determined by a timescale
on which the ionization is maintained.
A typical lifetime of ionizing stars is
$\lesssim 10$ Myr
(Bressan \textit{et al}., 1993; Inoue \textit{et al}., 2000).
We basically adopt $t=10$ Myr in this paper, but
we also consider longer durations for a longer
starburst duration or shattering over multiple starburst
episodes.}

\subsection{Parameters}\label{subsec:para}

Based on the discussions in the previous subsection,
we {change} the following parameters related to the
fragments.
The models are summarized in Table~\ref{tab:model},
where Model A adopts ``fiducial'' values for the
parameters.

\begin{table*}[t]
\renewcommand{\arraystretch}{1.2}
\vspace{-.3cm}
\caption{Models.}
\vspace{-.1cm}
\begin{center}
\begin{tabular}{lccccc} \hline
Name & species & $f_M$ & $P_1$ &
$\alpha_\mathrm{f}$ & grain velocities\\
 & & & (dyn cm$^{-2}$) & & \\ \hline
A$^a$ & silicate & 0.4 & $3\times 10^{11}$ & 3.3 & Yan \textit{et al}.\ (2004)\\
   & graphite & 0.4 & $4\times 10^{10}$ & 3.3 & Yan \textit{et al}.\ (2004)\\ \hline
B & silicate & 0.3 & $3\times 10^{11}$ & 3.3 & Yan \textit{et al}.\ (2004)\\
   & graphite & 0.3 & $4\times 10^{10}$ & 3.3 & Yan \textit{et al}.\ (2004)\\ \hline
C & silicate & 0.6 & $3\times 10^{11}$ & 3.3 & Yan \textit{et al}.\ (2004)\\
   & graphite & 0.6 & $4\times 10^{10}$ & 3.3 & Yan \textit{et al}.\ (2004)\\ \hline
D$^a$ & silicate & 0.4 & $1\times 10^{11}$ & 3.3 & Yan \textit{et al}.\ (2004)\\
   & graphite & 0.4 & $1.3\times 10^{10}$ & 3.3 & Yan \textit{et al}.\ (2004)\\ \hline
E$^a$ & silicate & 0.4 & $9\times 10^{11}$ & 3.3 & Yan \textit{et al}.\ (2004)\\
   & graphite & 0.4 & $1.2\times 10^{11}$ & 3.3 &  Yan \textit{et al}.\ (2004)\\ \hline
F & silicate & 0.4 & $3\times 10^{11}$ & 2.3 & Yan \textit{et al}.\ (2004)\\
   & graphite & 0.4 & $4\times 10^{10}$ & 2.3 & Yan \textit{et al}.\ (2004)\\ \hline
G & silicate & 0.4 & $3\times 10^{11}$ & 4.3 & Yan \textit{et al}.\ (2004)\\
   & graphite & 0.4 & $4\times 10^{10}$ & 4.3 & Yan \textit{et al}.\ (2004)\\ \hline
H & silicate & 0.4 & $3\times 10^{11}$ & 3.3 & Eq.\ (\ref{eq:vel})\\
   & graphite & 0.4 & $4\times 10^{10}$ & 3.3 & Eq.\ (\ref{eq:vel})\\ \hline
\end{tabular}
\\
$^a${We also examine KT10's formulation for fragments
(Section \ref{subsubsec:Pcr}).}
\end{center}
\label{tab:model}
\end{table*}

\subsubsection{Ejected fraction of shocked material}

How much fraction of the shocked material is finally
ejected as fragments is uncertain. This factor is
denoted as $f_M$ in Eq.~(\ref{eq:frag}).
According to Jones \textit{et al}.\ (1996),
$f_M$ is between 0.3 and 0.6, while H10 adopted 0.4.
Thus, we adopt 0.4 as a fiducial value (Model A) and
also examine 0.3 (Model B) and 0.6 (Model C).

\subsubsection{Critical pressure}
\label{subsubsec:Pcr}

Considering the uncertainty in the actual materials of
astronomical grains, we vary $P_1$ by
an order of magnitude as Models D and E in
Table \ref{tab:model}. The standard value adopted
from Jones \textit{et al}.\ (1996) is
$P_1=3\times 10^{11}$ and $4\times 10^{10}$
dyn cm$^{-2}$ for silicate and graphite, respectively
(Table \ref{tab:species}; Model A in Table \ref{tab:model}).
We adopt three times smaller and larger values
for $P_1$ in Models {D and E}, respectively.
We call grains with low/high $P_1$ soft/hard grains.

{There is another possible way of including the
critical pressure in the model based on Kobayashi
and Tanaka (2010, hereafter KT10).}
Jones \textit{et al}.\ (1996) constructed the fragmentation
model based on 
shock wave propagation caused by a collision (see also
Mizutani \textit{et al}., 1990). 
On the other hand, Holsapple (1987) provided a scaling
formula for collisional outcomes
based on laboratory experiments of his group
(the review of the scaling formula is seen in Holsapple, 1993). 
In this formula, $M_\mathrm{ej}$ is proportional to
\begin{eqnarray}
\varphi =\frac{E_\mathrm{imp}}{m_1Q_\mathrm{D}^\star}
\label{eq:phi}
\end{eqnarray}
for $\varphi\ll 1$, where
\begin{eqnarray}
E_\mathrm{imp}=\frac{1}{2}\frac{m_1m_2}{m_1+m_2}v^2
\label{eq:eimp}
\end{eqnarray}
is the impact energy between $m_1$ and $m_2$, and
$Q_\mathrm{D}^\star$ is the specific impact energy at
$M_\mathrm{ej}=m_1/2$. Since $M_\mathrm{ej}\sim m_1$
for $\varphi\gg 1$, KT10 simply
connected the cases for $\varphi\ll 1$ and $\varphi\gg 1$,
providing a simple formula for $M_\mathrm{ej}$ as
\begin{eqnarray}
M_\mathrm{ej}=\frac{\varphi}{1+\varphi}m_1.
\label{eq:frag_kob}
\end{eqnarray}

{Now we relate $Q_\mathrm{D}^\star$ to
$P_1$.}
The dominant channel for the small grain production
is cratering of large grains by small grains. If we
assume that $m_1\gg m_2$ (and as long as
$v^2/Q_\mathrm{D}^\star$ is not much larger than
unity), we obtain
$M_\mathrm{ej}\simeq m_2v^2/(2Q_\mathrm{D}^\star )$.
On the other hand, if we approximate Eq.\ (\ref{eq:frag})
as $M_\mathrm{ej}/m_2\simeq A(\rho_\mathrm{gr}v^2/P_1)$,
$A\sim 1$ for the range of quantities in this paper.
Therefore, if the two models are equivalent,
$Q_\mathrm{D}^\star\sim P_1/(2\rho_\mathrm{gr})$.

{Thus, we examine KT10's
formulation as follows. We adopt
Eq.\ (\ref{eq:frag_kob}) instead of Eq.\ (\ref{eq:frag})
for the total mass of fragments
with the same fragment size distribution
($\propto a^{-\alpha_\mathrm{f}}$ with the same
upper and lower grain radii). The value of $\varphi$
is evaluated by Eqs.\ (\ref{eq:phi}) and (\ref{eq:eimp}),
where $Q_\mathrm{D}^\star$ is given by}
\begin{eqnarray}
Q_\mathrm{D}^\star =\frac{P_1}{2\rho_\mathrm{gr}}.
\label{eq:qd}
\end{eqnarray}
{
This equation indicates that, if the energy per unit volume
given to the grain exceeds
$\sim P_1/\rho_\mathrm{gr}$, the major part of the grain
is disrupted.
We test Models A, D, and E by adopting KT10's
formulation.}

{Note that we have adopted the approximation,
$M_\mathrm{ej}\sim\rho_\mathrm{gr}v^2/P_1$.
This is a good approximation for $4s\phi_1>1$
($\phi_1=P_1/(\rho_\mathrm{gr}c_0^2$);
see Section \ref{subsec:shatter}).
However, if $\phi_1\ll 1$,
$M_\mathrm{ej}/m_2\sim (\rho_\mathrm{gr}v^2/P_1)/\phi_1$
for Eq.\ (\ref{eq:frag}), so that
$Q_\mathrm{D}^\star =(P_1/2\rho_\mathrm{gr})\phi_1$
gives a better approximation for $\phi_1\ll 1$.
In other words, if $\phi_1$ is significantly smaller
than 1 (i.e., for soft dust), Eq.\ (\ref{eq:qd})
overestimates $Q_\mathrm{D}^\star$ by a factor of $\phi_1$.
Thus, if we calculate shattering
by adopting  Eq.\ (\ref{eq:qd}), grains are shattered
less compared with Jones \textit{et al}.'s case for soft
dust. In this paper, because it is easy to interpret
Eq.\ (\ref{eq:qd}), we simply adopt it for KT10's
formulation.}

\subsubsection{Size distribution of shattered fragments}
\label{subsubsec:frag}

The power index of the size distribution of shattered
fragments, $\alpha_\mathrm{f}$, may be
reflected in the grain size distribution after
shattering. The canonical value adopted in
H10 is $\alpha_\mathrm{f}=3.3$, which is based
on Jones \textit{et al}.\ (1996). Here we
also examine the cases of $\alpha_\mathrm{f}=2.3$
and 4.3 {(Models F and G, respectively)},
considering a wide range obtained in
experiments (Takasawa \textit{et al}., 2011).

\subsubsection{Grain velocities}
\label{subsubsec:vel}

\begin{figure*}[t]
\includegraphics[width=0.48\textwidth]{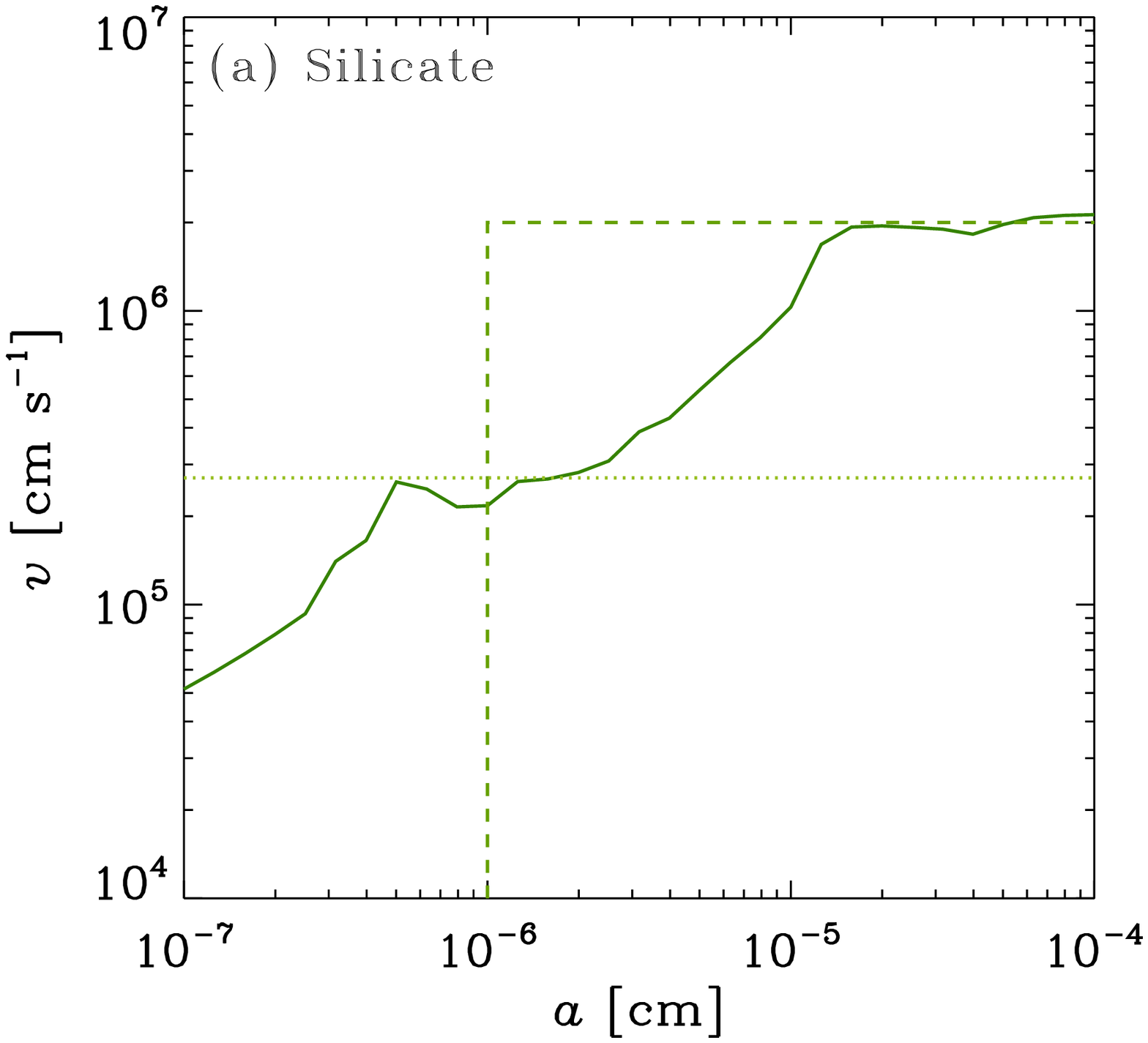}
\includegraphics[width=0.48\textwidth]{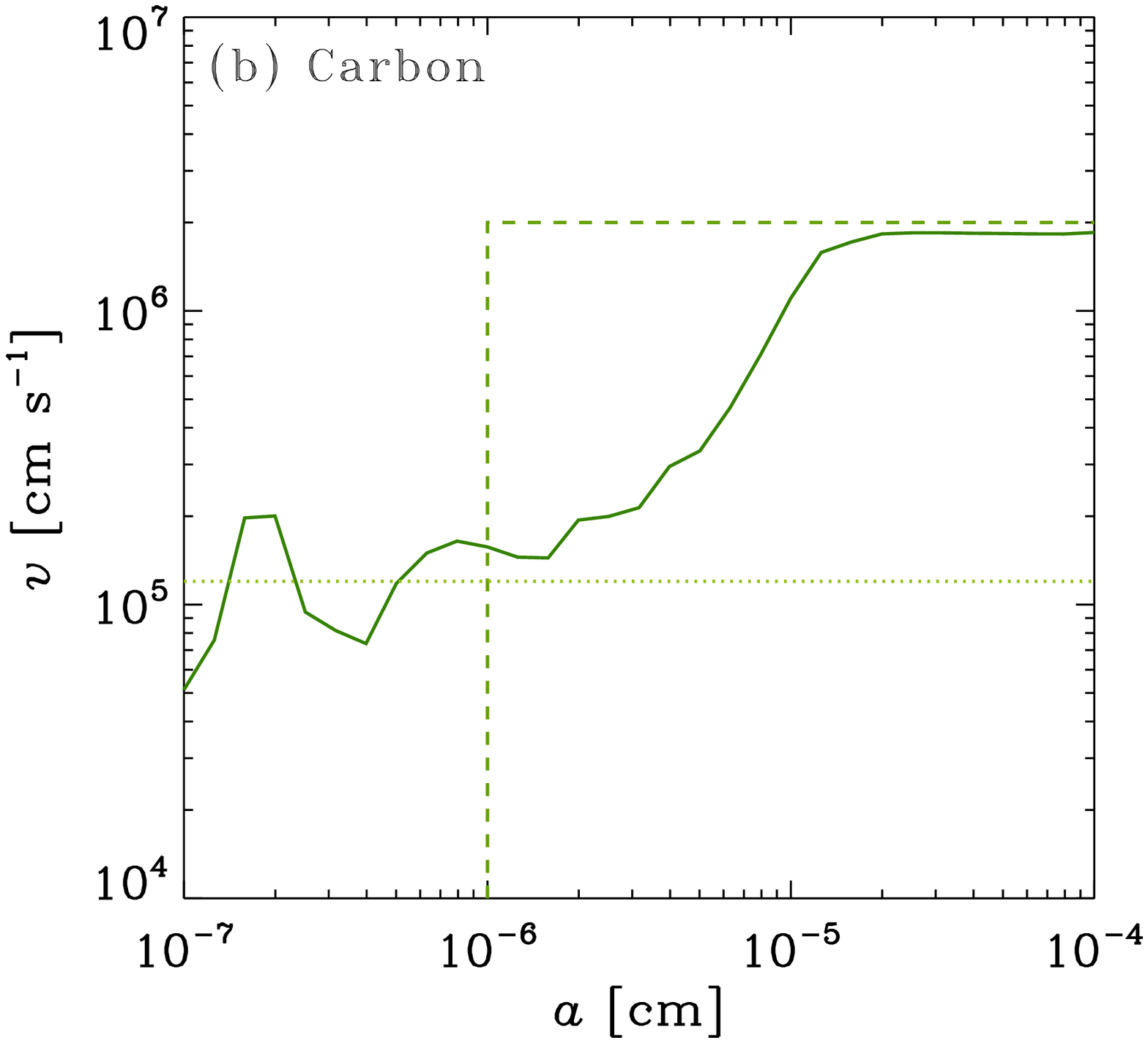}
\caption{\footnotesize
Grain velocities $v$ as a function of grain radius $a$. Two
grain species, (a) silicate and (b) carbonaceous dust,
are shown. The solid lines indicate the velocity calculated
by Yan \textit{et al}.\ (2004) with the physical condition of
the WIM described in the text. The dashed line shows
the simplified velocity model (Eq.\ \ref{eq:vel}) with
$a_\mathrm{c}=10^{-6}$ cm. The dotted line shows the
shattering threshold velocity.
}
\label{fig:velocity}
\end{figure*}

Grain motions are mainly driven by gyroresonance
in the diffuse medium (Yan \textit{et al}., 2004).
Fig.\ \ref{fig:velocity} presents the grain velocities
as a function of grain radius in the WIM
(Section \ref{subsec:shatter}). Large grains with
$a\gtrsim 10^{-5}$ cm are accelerated to
$\sim 20$ km s$^{-1}$ (i.e., similar
velocities to the Alfv\'{e}n velocity).
Since larger grains are coupled with larger-scale
turbulence, the motions of larger
grains are less affected by the damping by
dissipation, which occurs on small scales.
The dissipation scale depends on the physical
condition of the ISM, especially on the ionization
degree. Thus, the critical grain size, above
which the grains are efficiently accelerated by
gyroresonance, varies by the physical condition
of the ISM and the power spectrum of the turbulence.
Because of such a variation of the critical grain size,
it is worth investigating how the shattered grain
size distribution changes as a result of the
variation of the critical grain size.

To make clear the dependence of the critical grain
size denoted as $a_\mathrm{c}$, we also examine
a model in which the
size dependence of the grain velocity dispersion is
simplified as
\begin{eqnarray}
v(a)=\left\{
\begin{array}{cc}
v_\mathrm{gyro} & \mbox{if $a\ge a_\mathrm{c}$},\\
0 & \mbox{if $a<a_\mathrm{c}$}.
\end{array}
\right.\label{eq:vel}
\end{eqnarray}
An example with $a_\mathrm{c}=10^{-6}$ cm is
shown in Fig.\ \ref{fig:velocity}.
We fix $v_\mathrm{gyro}=20$ km s$^{-1}$. As long as
$v_\mathrm{gyro}>v_\mathrm{sh}$, the variation of
$v_\mathrm{gyro}$ changes the timescale of
shattering, which is roughly scaled with
$v_\mathrm{gyro}^{-1}$.
We examine $a_\mathrm{c}=10^{-5}$ and $10^{-6}$ cm.
{This treatment of velocities is labeled as
Model H.}

\section{Results}\label{sec:result}

\subsection{Ejected fraction of shocked material}

\begin{figure*}[t]
\includegraphics[width=0.48\textwidth]{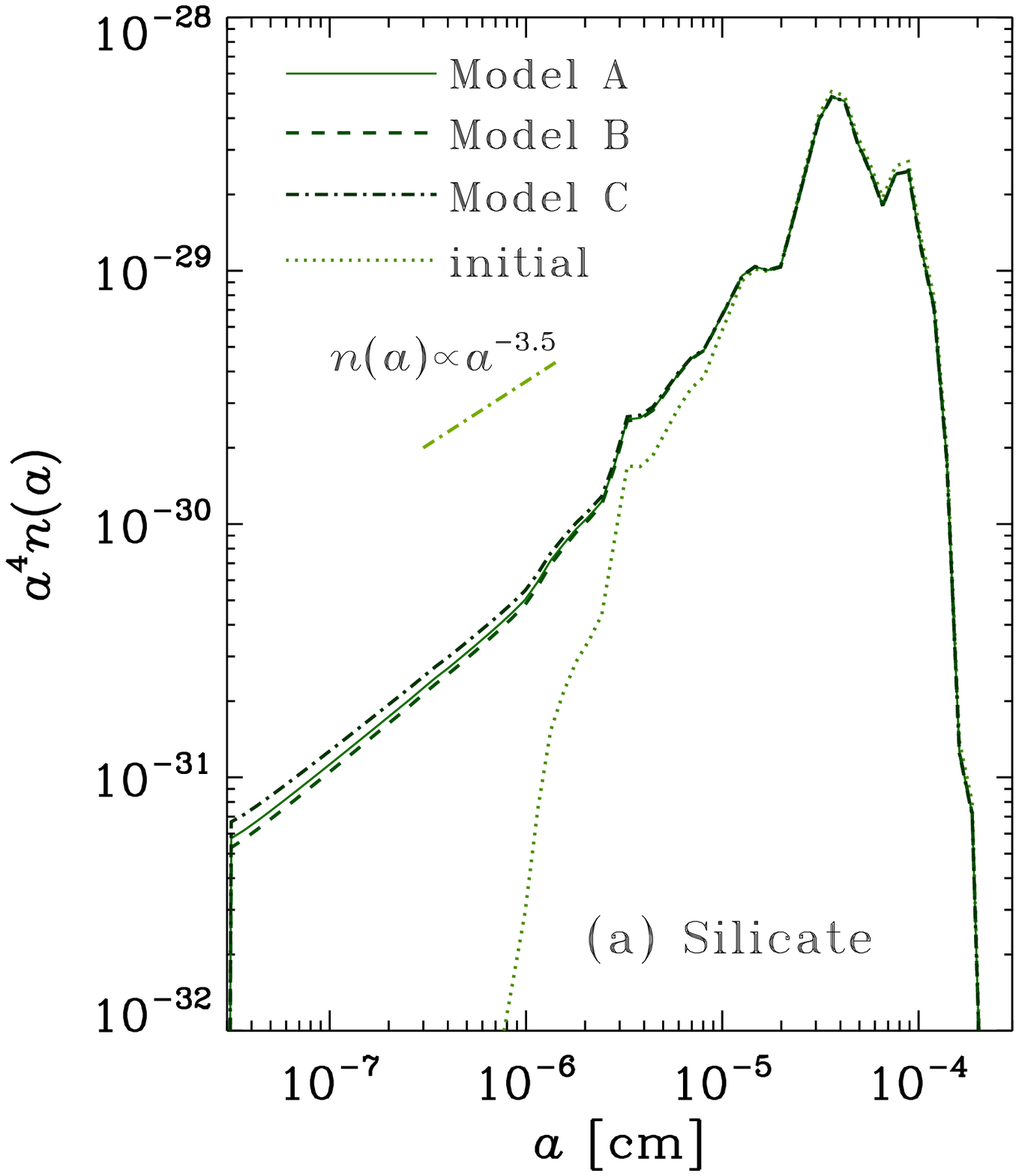}
\includegraphics[width=0.48\textwidth]{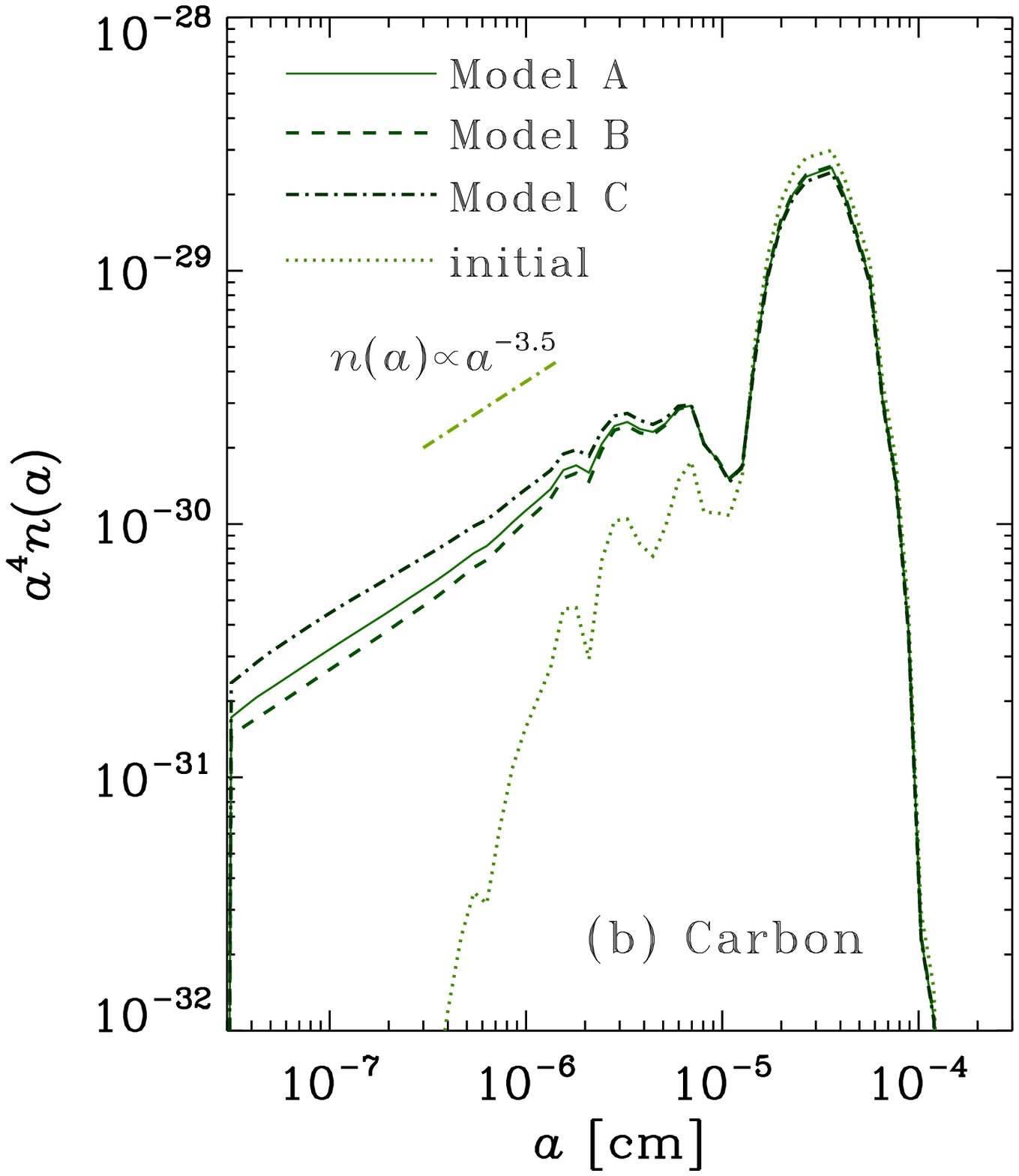}
\caption{\footnotesize
Grain size distributions presented by multiplying
$a^4$ to show the mass distribution in each logarithmic bin of
the grain radius.
The solid, dashed, and dot-dashed lines show the results
at $t=10$ Myr for Models A, B, and C ($f_M=0.4$, 0.3, and
0.6), respectively.
The dotted line presents the initial
grain size distribution before shattering.
Two grain species,
(a) silicate and (b) carbonaceous dust, are shown.
We also show a slope of 0.5 [i.e., $n(a)\propto a^{-3.5}$].
}
\label{fig:fM}
\end{figure*}

We examine the effect of $f_M$ on the grain size distribution.
In Fig.\ \ref{fig:fM}, we show the grain size distributions at
$t=10$~Myr for Models A, B, and C. In this paper,
the grain size distributions are presented by multiplying
$a^4$ to show the mass distribution in each logarithmic bin of
the grain radius. We observe that the production of small grains
is the most efficient for the largest $f_M$. This is simply
because a larger fraction of the shocked material is ejected
as fragments. However, the abundance of the
smallest grains is not simply proportional to $f_M$,
since the catastrophic disruption does not depend on
$f_M$. The maximum ratio of $n(a)$ between
$f_M=0.3$ and 0.6 is 1.3 for silicate and 1.6 for carbonaceous
dust.

\subsection{Critical pressure}

\begin{figure*}[t]
\includegraphics[width=0.48\textwidth]{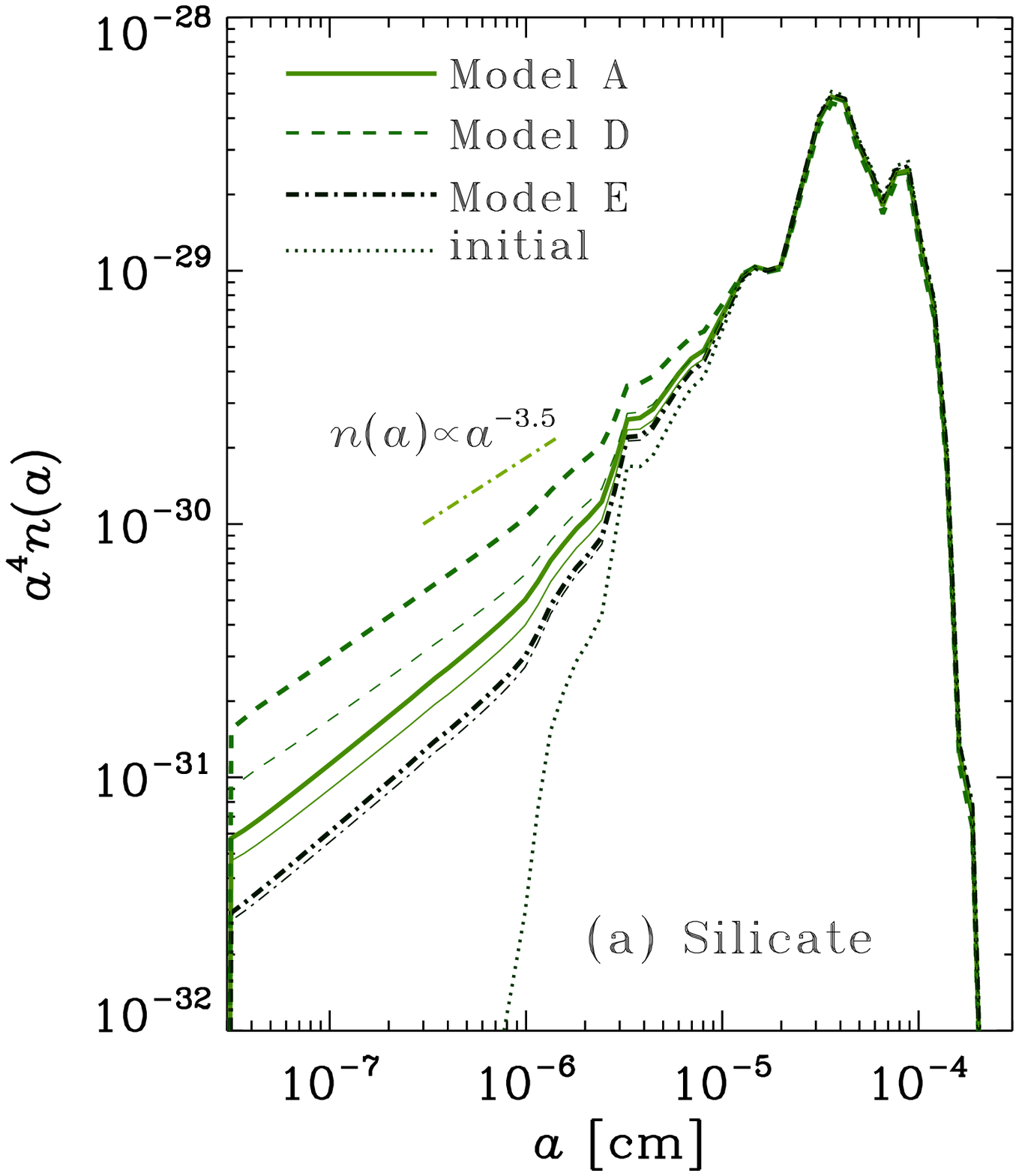}
\includegraphics[width=0.48\textwidth]{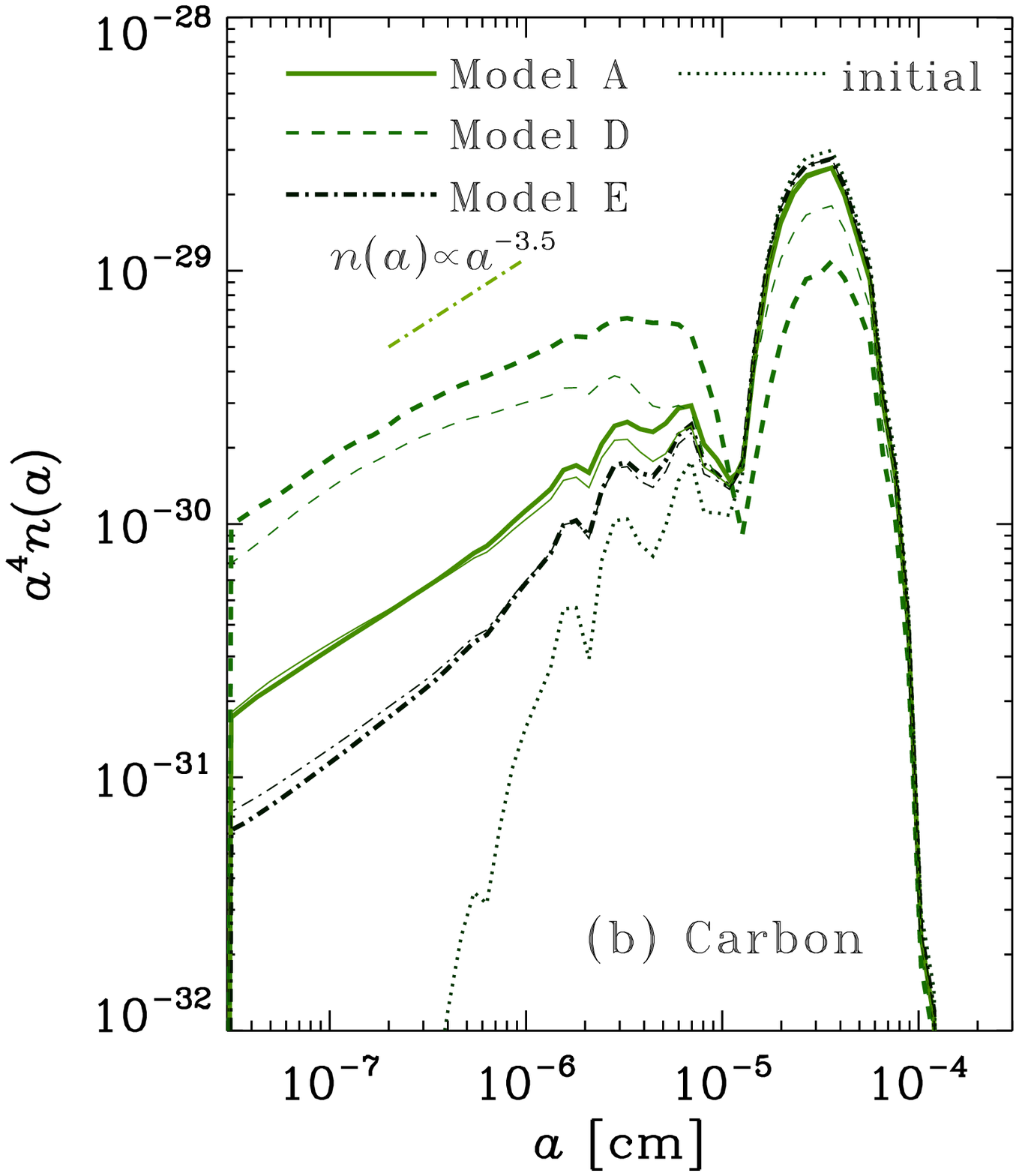}
\caption{\footnotesize
Same as Fig.\ \ref{fig:fM} but for Models A, D, and E.
The thick solid, dashed, and dot-dashed lines show the results
at $t=10$ Myr for Models A, D, and E (fiducial, hard, and soft
grains),
respectively, while the thin lines adopt the
treatment of fragments by KT10
(the same line species shows the same model).
The dotted line presents the initial
grain size distribution before shattering.
Two grain species,
(a) silicate and (b) carbonaceous dust, are shown.
}
\label{fig:Pcr}
\end{figure*}

We examine the dependence on the critical pressure
($P_1$). The adopted values for the fiducial case (Model A)
is shown in Table \ref{tab:model},
while smaller and larger values for $P_1$ are also
examined (Models D and E, respectively).
In Fig.\ \ref{fig:Pcr}, we compare Models A,  D, and E
at $t=10$ Myr.
We find that the effect of $P_1$ is significant.
{Carbonaceous dust in Model D} has the lowest
$P_1$, so that the largest grains
are efficiently shattered into smaller sizes.
The mass loss rate of large grains with mass $m$ 
is {proportional to $P_1^{-8/9}$ (Section \ref{subsec:shatter})};
hence a low $P_1$ increases the number density of small
($a\lesssim 10^{-6}$ cm) grains. 
On the other hand, collisional cascades determine the slope of 
the grain size distribution as $-3.5$, independent of $P_1$
(Hellyer, 1970; Tanaka \textit{et al}., 1996; KT10). 
The slope is therefore given by $-3.5$ for $a\lesssim 10^{-6}$ cm,
a value ($-3.5$) consistent with
one derived observationally for the dust grains in the Milky Way
and the Magellanic Clouds
(Mathis \textit{et al}., 1977; Pei, 1992).

To show the dependence on the treatment of total
cratering volume, we also show the results calculated
by the simple formulation of KT10
(see Section \ref{subsubsec:Pcr} for the formulation)
{in Fig.\ \ref{fig:Pcr}}. We observe that
KT10's formulation
is in fair agreement with our formulation
based on Jones \textit{et al}.\ (1996)
for the fiducial (Model A) and hard dust
{(Model E)},
but the discrepancy is relatively large for the soft
dust {(Model D) because
Eq.\ (\ref{eq:qd}) is not a good approximation
for the soft dust (see Section \ref{subsubsec:Pcr}
for detailed discussions).}
The maximum ratio between the two formulations
in {Model D} is
1.8 for silicate and 2.2 for carbonaceous dust.
For carbonaceous dust, the maximum difference occurs
around the complicated feature around $a\sim 10^{-5}$ cm,
but if we focus on the small grain production at
$a<10^{-6}$ cm, the ratio is 1.4.
Thus, the uncertainty caused by the different
formulations is just comparable to that
caused by the difference in $f_M$. If we adopt the
standard values for $P_1$ (i.e., Model~A), it is
concluded that the small grain production by shattering
is described by the simple formulation by KT10 as
precisely as more complicated
framework of Jones \textit{et al}.\ (1996).

\subsection{Size distribution of shattered fragments}
\label{subsec:frag}

\begin{figure*}[t]
\includegraphics[width=0.48\textwidth]{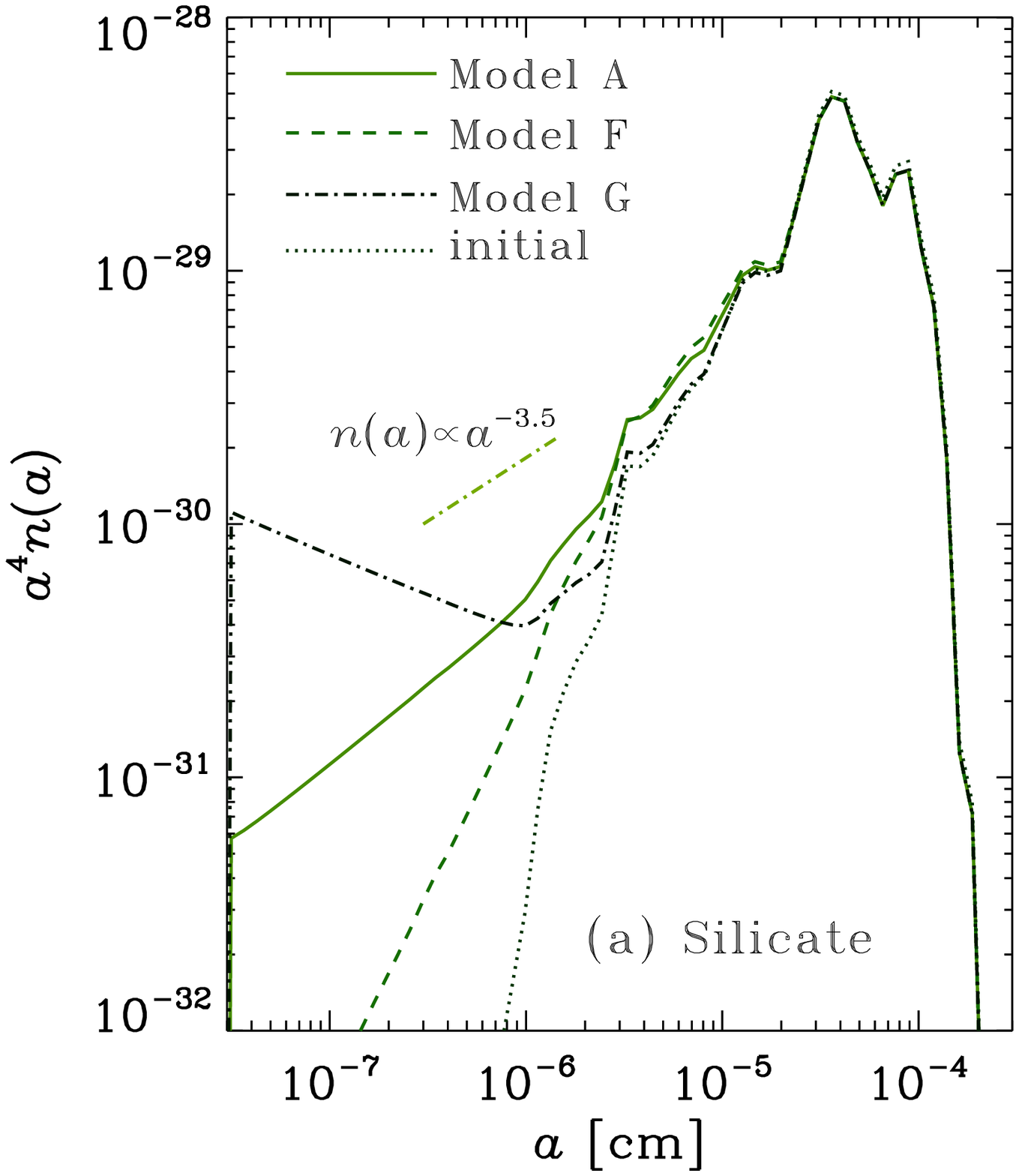}
\includegraphics[width=0.48\textwidth]{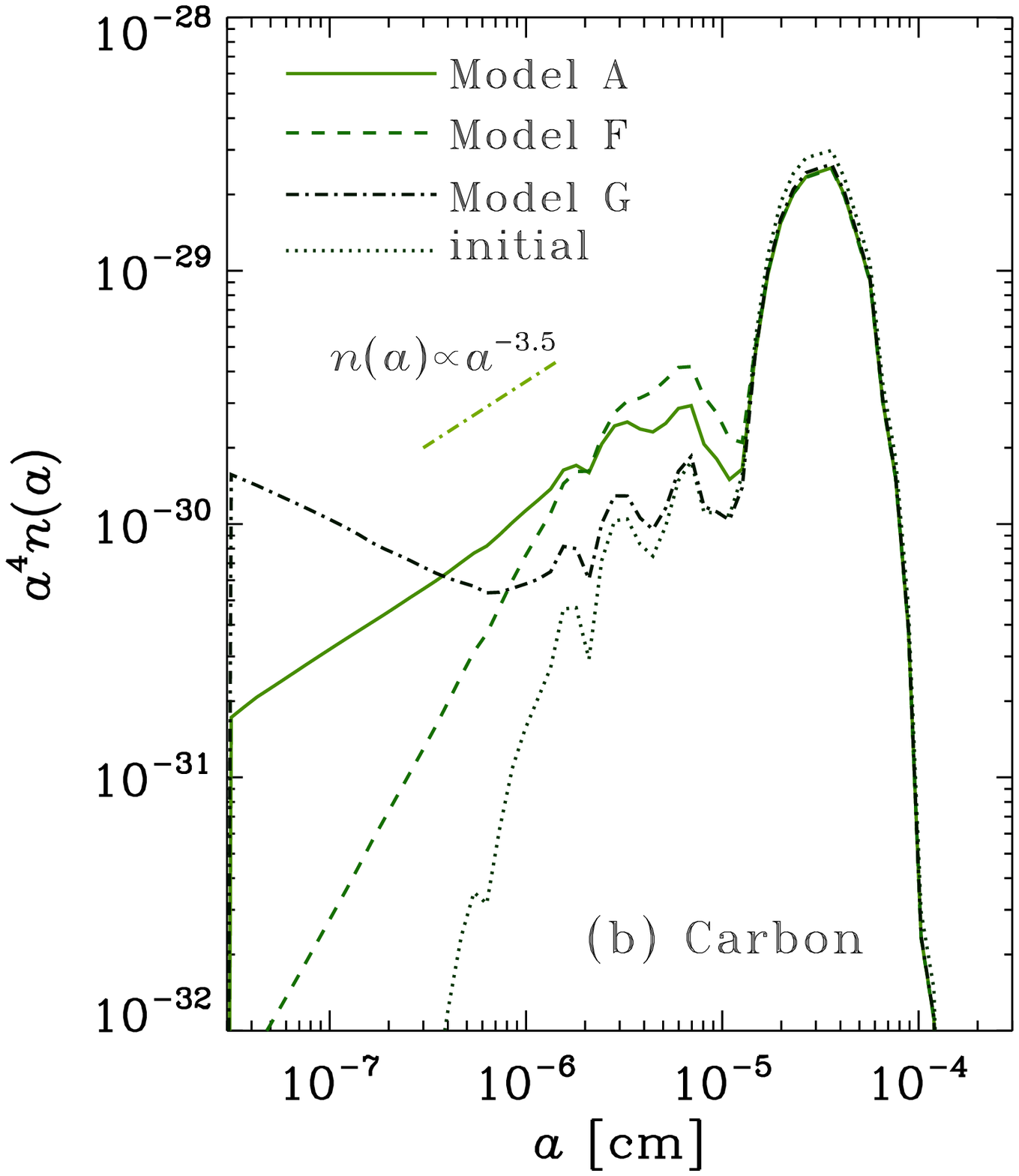}
\caption{\footnotesize
Same as Fig.\ \ref{fig:fM} but for Models A, F, and G,
in order to investigate the effect of $\alpha_\mathrm{f}$
(power index of the grain size distribution of shattered
fragments). The solid, dashed, and dot-dashed lines show
the results at $t=10$~Myr for Models A, D, and E
($\alpha_\mathrm{f}=3.3$, 2.3, and 4.3),
respectively. The dotted line presents the initial
grain size distribution before shattering.
}
\label{fig:alpha}
\end{figure*}

As we mentioned in Section \ref{subsubsec:frag},
we assume that the
size distribution of shattered fragments can be
described by a power law function with
power index $\alpha_\mathrm{f}$.
In Fig.\ \ref{fig:alpha}, we show the results for
different values of $\alpha_\mathrm{f}$ (Models A,
F, and G). We observe that the difference in
$\alpha_\mathrm{f}$ affects the resulting
grain size distributions. However,
we also see that the difference between
$\alpha_\mathrm{f}=3.3$ and 2.3 (Models A and F)
is relatively small especially for silicate
at $a>10^{-6}$ cm, which
implies that the grain size distribution does
not necessarily becomes
$n(a)\propto a^{-\alpha_\mathrm{f}}$.
The grain size distribution in Model G at
$a>10^{-6}$ cm is near to the initial grain size
distribution, since the shattered grains are
predominantly distributed at the smallest
($a\ll 10^{-6}$ cm) sizes.

\begin{figure*}[t]
\includegraphics[width=0.48\textwidth]{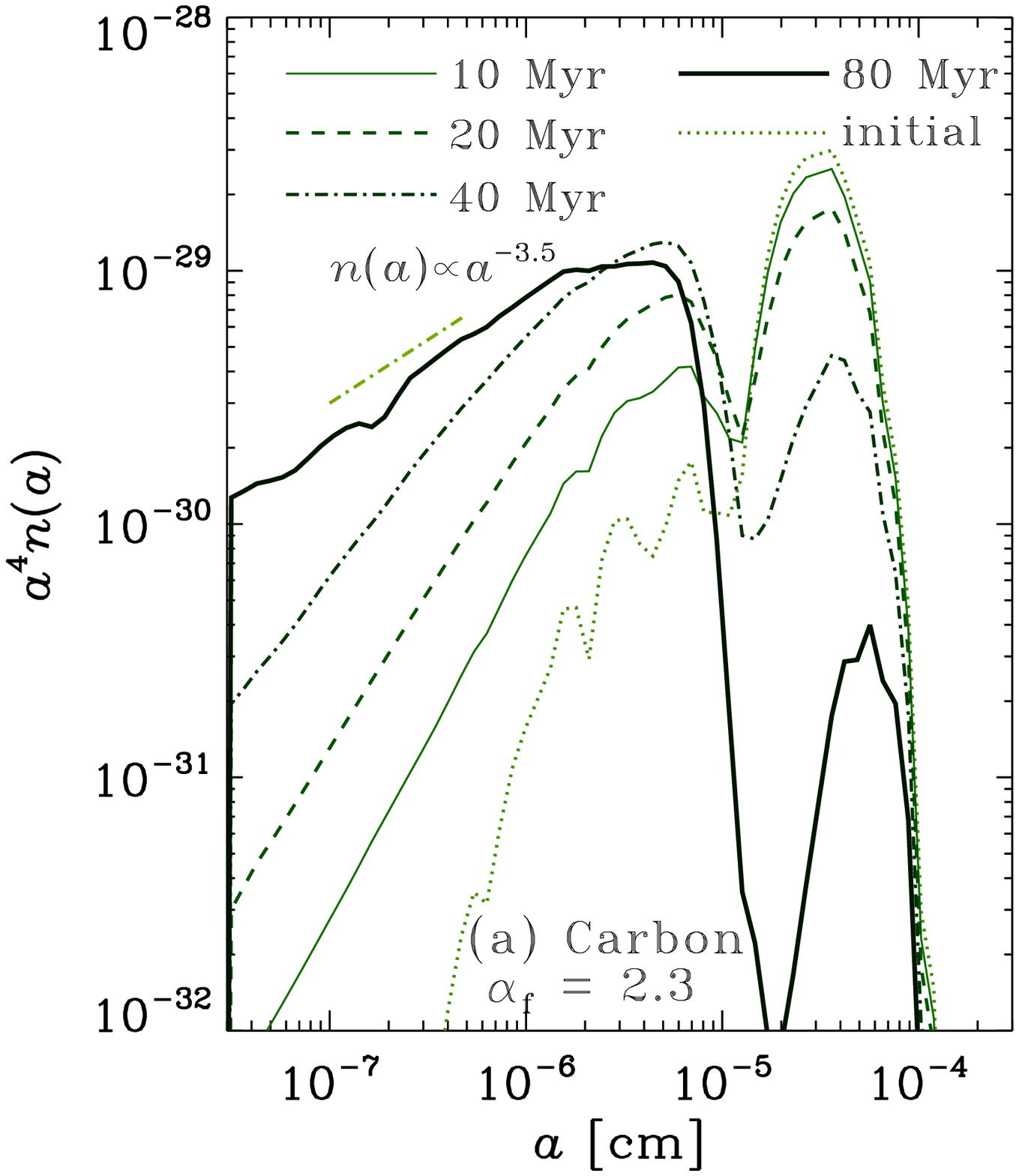}
\includegraphics[width=0.48\textwidth]{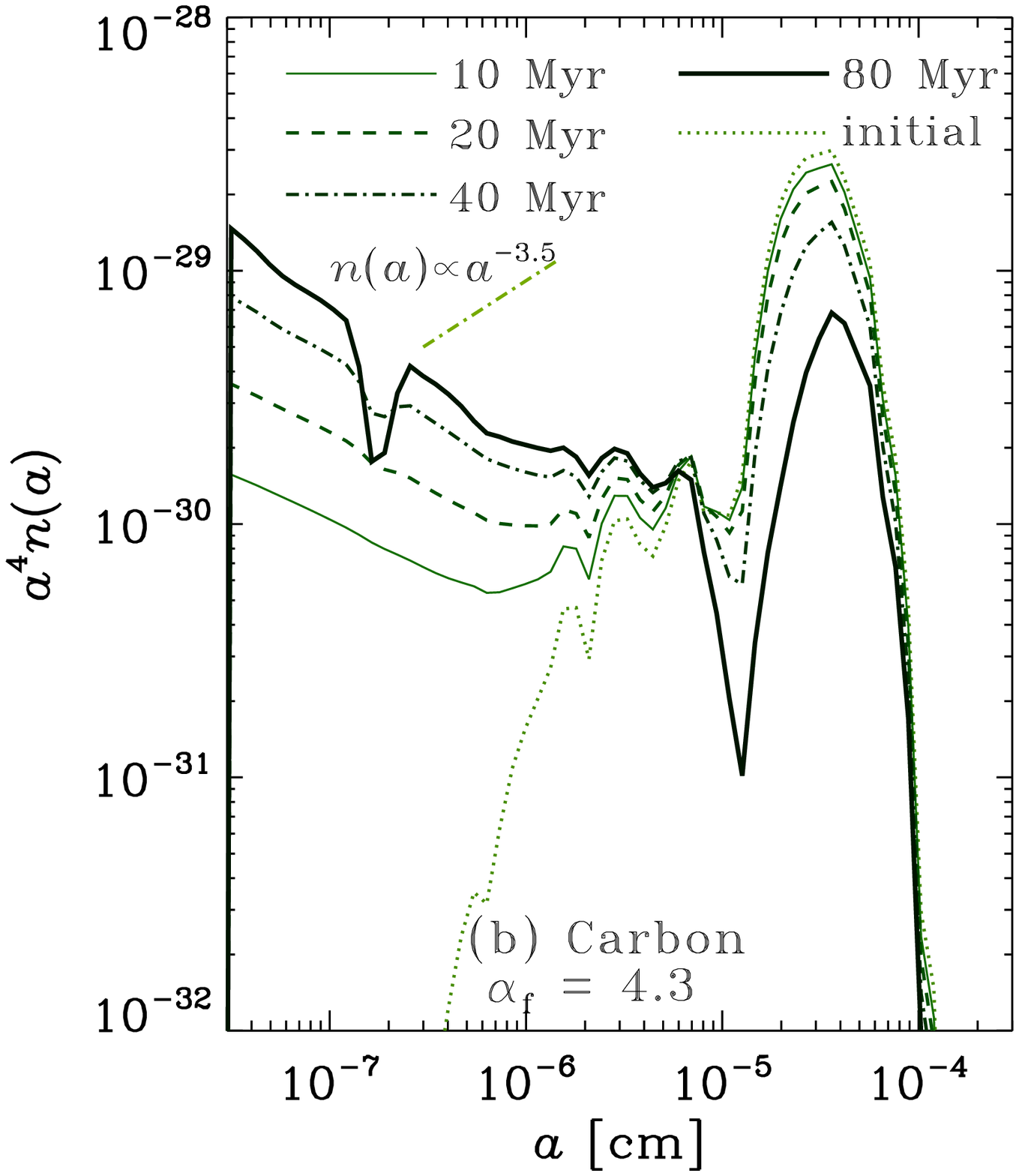}
\caption{\footnotesize
Time evolution of grain size distribution for Models F and G
in Panels (a) and (b), respectively, for carbonaceous dust.
The thin solid, dashed, dot-dashed, and thick solid lines
represent $t=10$, 20, 40, and 80 Myr, respectively.
The dotted line presents the initial grain size distribution before
shattering.
}
\label{fig:alpha_ev}
\end{figure*}

In order to examine more clearly how the grain size
distribution evolves depending on $\alpha_\mathrm{f}$,
we show the results of Models F and G for
$t=10$, 20, 40, and 80 Myr in Fig.\ \ref{fig:alpha_ev}.
The dip around $a\sim 2\times 10^{-7}$ cm, which is
clearly seen at $t=80$ Myr, is because of the velocity
exceeding the shattering threshold around this grain
radius (Fig.\ \ref{fig:velocity}).
This non-monotonic behavior of the velocity is
due to the complexity of the grain charge as a function
of grain size (Yan \textit{et al}., 2004).

There is a qualitative difference in the evolution of
grain size distribution between $\alpha_\mathrm{f}=2.3$
and 4.3. For $\alpha_\mathrm{f}=2.3$,
the slope of the grain size distribution at
$a\lesssim 10^{-6}$ cm changes, while
for $\alpha_\mathrm{f}=4.3$, it evolves with a constant
slope [$n(a)\propto a^{-\alpha_\mathrm{f}}$] at
small sizes. In the former case, the slope of
the grain size distribution seems to approach
$n(a)\propto a^{-3.5}$. The variation of slope occurs in a
top-down manner; for example, for carbonaceous dust,
the grain size distribution is nearly
$n(a)\propto a^{-3.5}$ between $a=10^{-6}$ and
$10^{-5}$ cm at 40 Myr, while it approaches
$n(a)\propto a^{-3.5}$ at smaller sizes later at 80 Myr.
This top-down behavior is because of the nature of
shattering, which disrupts large grains into
a lot of small pieces. This trend is also seen for
$\alpha_\mathrm{f}=3.3$. Generally,
if $\alpha_\mathrm{f}\lesssim 3.5$,
the slope of grain size distribution tends to approach
the MRN value ($-3.5$). On the contrary,
if $\alpha_\mathrm{f}\gtrsim 4$ (see also
Section \ref{subsec:sensitivity}),
the resulting grain size distribution is determined
by the grain size distribution of shattered fragments.
Indeed, if $\alpha_\mathrm{f}>4$, the smallest
particles, which are too small to disrupt
large grains around the peak (i.e., $a\gtrsim 10^{-5}$ cm),
have the dominant contribution to the
total fragment mass. Moreover, we remove the grains
reaching the smallest-size bin ($a=3\times 10^{-8}$ cm), and these
grains do not affect the subsequent evolution of
grain size distribution. Thus, for $\alpha_\mathrm{f}>4$,
the fragments just keep their grain size
distribution and eventually lost in reaching the smallest
grain size.

We note that continuous shattering for such a
long time as 80 Myr is not applicable to the real
WIM, since the lifetime of the WIM is probably
shorter than 80 Myr. Yet, the above calculations for
the long shattering durations show that,
if grains are repeatedly shattered with
$\alpha_\mathrm{f}\lesssim 3.5$, the final
grain size distribution can approach a power law
with an index of $\sim -3.5$. Therefore,
shattering is a promising mechanism to produce
an MRN-like grain size distribution. This issue
is further discussed in Section \ref{subsec:sensitivity}.

\subsection{Grain velocities}

\begin{figure*}[t]
\includegraphics[width=0.48\textwidth]{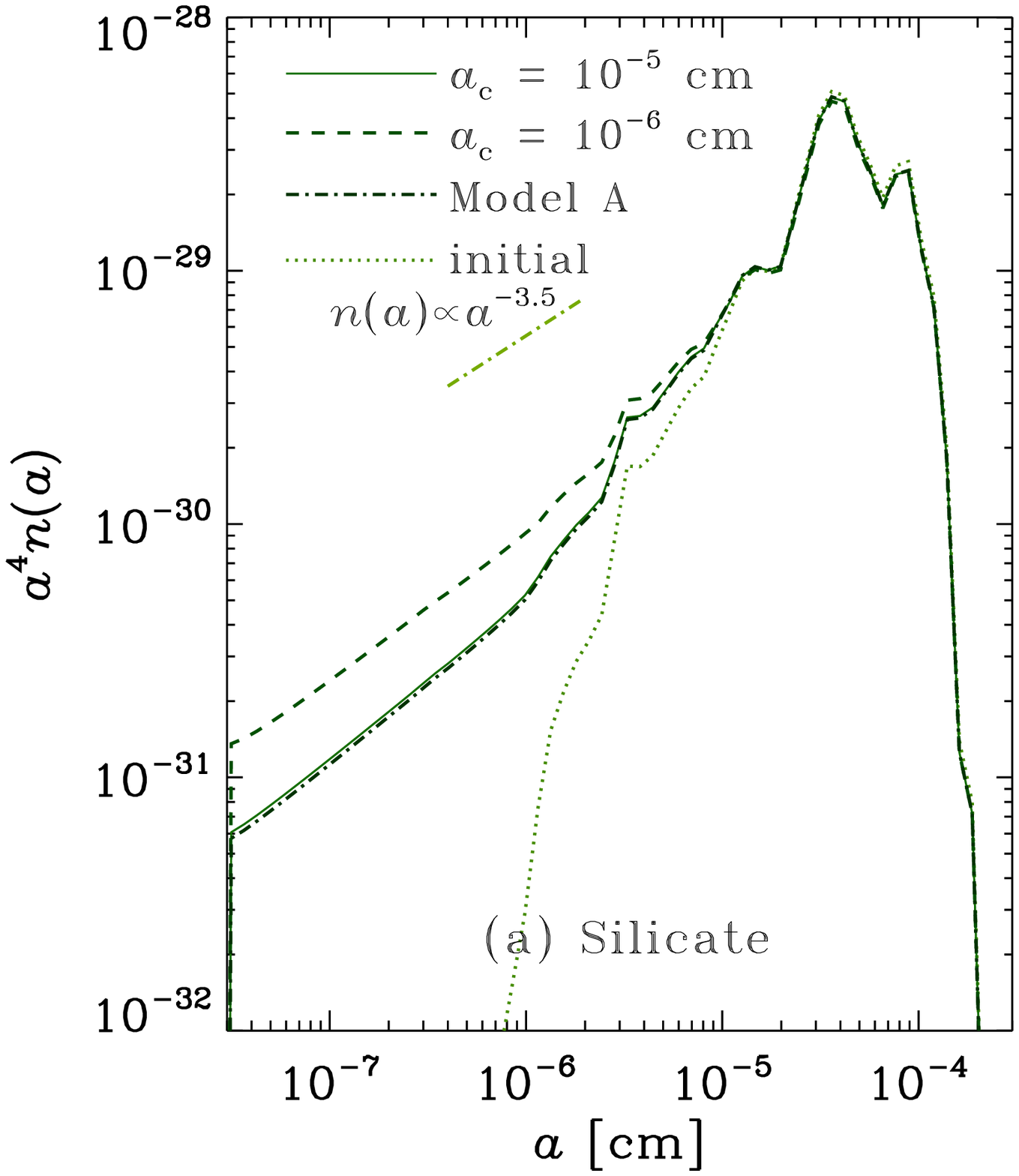}
\includegraphics[width=0.48\textwidth]{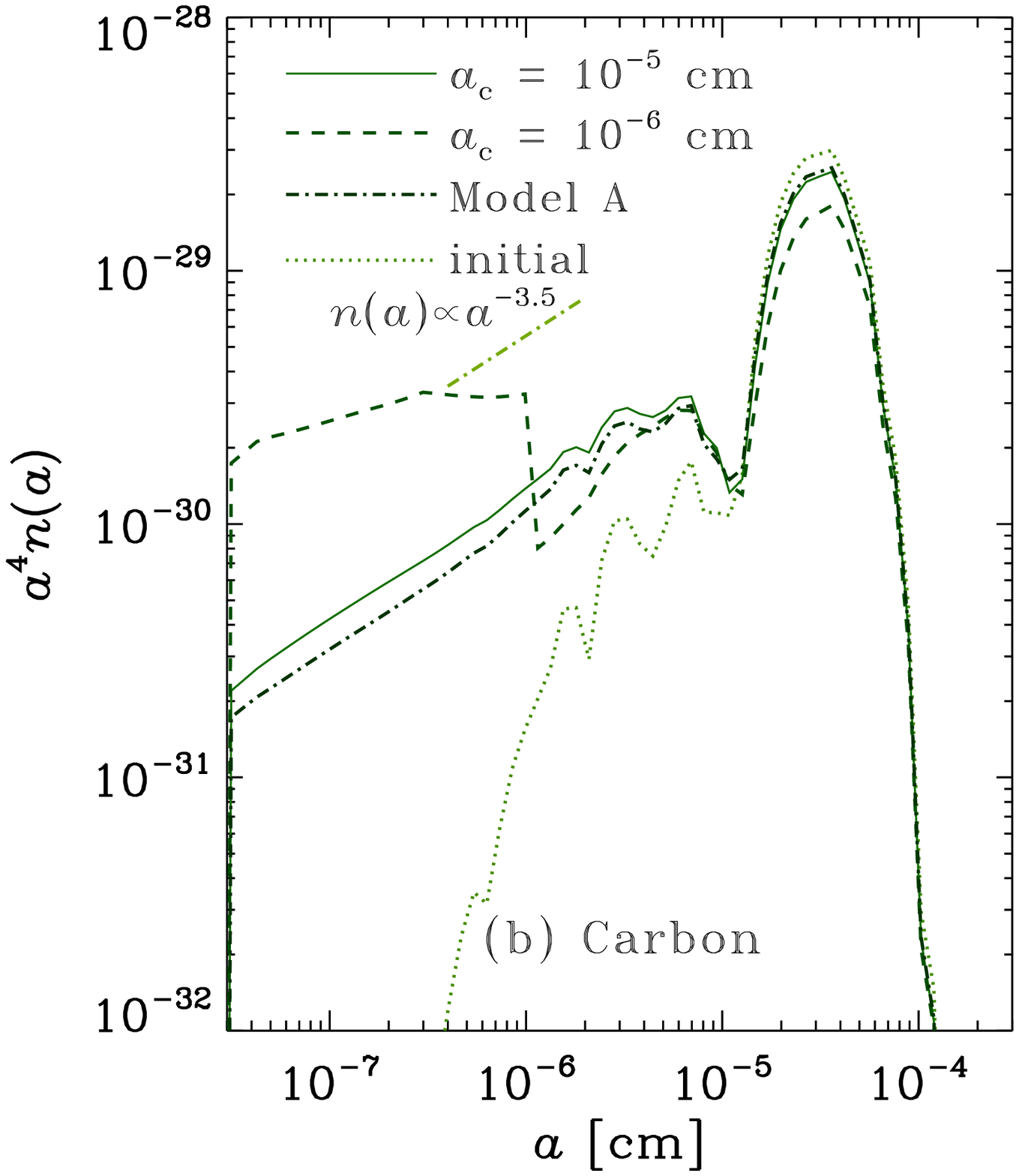}
\caption{\footnotesize
Same as Fig.\ \ref{fig:fM} but for Model H with various
$a_\mathrm{c}$.
The solid, and dashed lines show the results
for $a_\mathrm{c}=10^{-5}$ and $10^{-6}$ cm,
respectively. The dotted line presents the initial
grain size distribution before shattering, {and
the dot-dashed line shows the results for Model A
as a reference.}
}
\label{fig:vel}
\end{figure*}

To understand the effect of grain velocities,
we examine Model H, in which we use Eq.\ (\ref{eq:vel})
and vary $a_\mathrm{c}$. In Fig.~\ref{fig:vel},
we show the grain size distributions at $t=10$ Myr
for $a_\mathrm{c}=10^{-5}$ and $10^{-6}$ cm.
We observe that the difference is
clear at small grain sizes: because small grains
are also shattered efficiently for small $a_\mathrm{c}$,
the abundance of smallest-sized grains is larger
for smaller $a_\mathrm{c}$. Since carbonaceous dust
is softer than silicate, the effect of $a_\mathrm{c}$
has a larger imprint on the grain size distribution of
carbonaceous dust. Indeed, we clearly see the effect of
$a_\mathrm{c}$ in the grain size distribution of
carbonaceous dust in Fig.\ \ref{fig:vel} (right panel).
Thus, if small ($a\lesssim 10^{-6}$ cm) grains also
acquire a larger velocity than the shattering threshold,
it is predicted that the grain size distribution may
not be approximated by a simple power law.

{In Fig.\ \ref{fig:vel}, we also show the
results of Model A as a reference. Since the grain
velocities at $a<10^{-5}$ cm drops below
20 km s$^{-1}$ for Model A (Fig.\ \ref{fig:vel}),
Model H with $a_\mathrm{c}=10^{-5}$ cm
produce a more similar
grain size distribution to Model A than
Model H with $a_\mathrm{c}=10^{-6}$ cm.
Thus, the effect of $a_\mathrm{c}$ (i.e.,
the minimum size of grain size accelerated to
the maximum velocity) is more important than the
detailed functional form of $v(a)$ at $a<a_\mathrm{c}$.}

\section{Discussion}\label{sec:discussion}

\subsection{Timescales }\label{subsec:timescale}

{In order to interpret the results above,}
it is useful to estimate a typical timescale
of grain--grain collision. A grain with
$a=a_1$ collide with a grain whose typical
radius is $a_2$ with a frequency of
\begin{eqnarray}
\tau_\mathrm{coll}\sim
\frac{1}{\pi (a_1+a_2)^2n(a_2)\,\Delta a_2v_{12}},
\end{eqnarray}
{where $\Delta a_2$ is the radius interval of interest.
Note that
$n(a_2)\,\Delta a_2$ has a dimension of number
density (Section \ref{sec:model}). By using logarithmic
size interval $\Delta\log a_2$,
$\Delta a_2\sim a_2\Delta\log a_2$. If we consider
a logarithmic size bin so that $\Delta\log a_2$ is of
order unity, we can replace
$\Delta a_2$ with $a_2$.}
We estimate the typical timescale of small grain
production by adopting the timescale on which
a large grain is hit by a small
grain ($a_1\gg a_2$). In such a case,
we obtain
\begin{eqnarray}
\tau_\mathrm{coll} & \sim &
\frac{1}{\pi a_1^2n(a_2)a_2v_1}\nonumber\\
& \sim &
\displaystyle{
\frac{1}{\pi\left({\displaystyle\frac{a_1}{a_2}}\right)^2
[n(a_2)a_2^4]{\displaystyle\frac{1}{a_2}}v_1}
}.\nonumber\\
& \sim &
5.0\left(\frac{a_2}{a_1}\right)^2\left(
\frac{a_2}{10^{-6}~\mathrm{cm}}\right)\left[
\frac{n(a_2)a_2^4}{10^{-30}}\right]^{-1}\nonumber\\
& & \times\left(
\frac{v_1}{20~\mathrm{km~s}^{-1}}\right)^{-1}~\mathrm{Gyr}.
\end{eqnarray}
Thus, we find that the timescale of small grain
production becomes shorter if the typical size
of colliding grains ($a_2$) becomes small and
the abundance of such grain [$n(a_2)$] increases.
For the initial grain size distribution, the grain
radius is concentrated
in a relatively narrow range around
$\sim 3\times 10^{-5}$ cm
{(Section \ref{subsec:initial};
Fig.\ \ref{fig:fM} and subsequent figures also
show the initial grain
size distribution)}, so that
$a_2/a_1\sim 1/10$ and $a_2\sim 3\times
10^{-6}$ cm. Under such a condition,
$\tau_\mathrm{coll}\sim 1.5\times 10^8$ yr,
{which is long compared with the typical duration
treated in this paper (10 Myr).}
Thus, the large grains are hardly affected by
shattering which is consistent with the fact that
the grain size distribution around $10^{-5}$ cm is
intact at 10 Myr (Fig.\ \ref{fig:fM}).
However, after a small amount of
shattering, a lot of small grains are produced,
and if grains with $a\lesssim 10^{-6}$ cm achieve
$n(a)a^4\sim 10^{-30}$, the collision timescale
with such small grains
for a large grain ($a_1\sim 3\times 10^{-5}$ cm)
becomes an order of 10 Myr.

The timescale on
which a large grain is lost because of repeated
collisions of small grains can be estimated by
\begin{eqnarray}
\tau_\mathrm{disr}\sim (m_1/M_\mathrm{ej})
\tau_\mathrm{coll}.\label{eq:disr}
\end{eqnarray}
This timescale is called disruption timescale.
{
Here we show that the timescale of the loss of large
($a\sim 3\times 10^{-5}$ cm) grains is consistent
with the disruption timescale. For analytical
convenience, we adopt KT10's formulation, which is
roughly equivalent with Jones \textit{et al}.\ (1996)'s
formulation through Eq.\ (\ref{eq:qd}). Since we are interested in
the disruption of large grains by collisions with small grains,
$m_1\gg m_2$ holds. Thus, from Eqs.\ (\ref{eq:phi}) and (\ref{eq:qd}),
we obtain $\varphi\sim (a_2/a_1)^3(\rho_\mathrm{gr}v^2/P_1)$.
If we adopt $a_1\sim 3\times 10^{-5}$ cm,
$a_2\sim 3\times 10^{-6}$ cm, and $v\sim 20$ km s$^{-1}$,
we obtain $\varphi\sim 0.22$ for carbonaceous grains.
Thus, $M_\mathrm{ej}/m_1\sim 0.18$ from Eq.\ (\ref{eq:frag_kob}),
and we finally obtain $\tau_\mathrm{disr}\sim 5.5\tau_\mathrm{coll}$.
Recalling that $\tau_\mathrm{coll}$ is of an order of
10 Myr, the disruption timescale for the large
grains is a few tens of Myr. This timescale is
consistent with the loss of the grains around
$a\sim 3\times 10^{-5}$ cm in a few tens of Myr.
}

\subsection{Sensitivity to the parameters}
\label{subsec:sensitivity}

\begin{figure*}[t]
\includegraphics[width=0.48\textwidth]{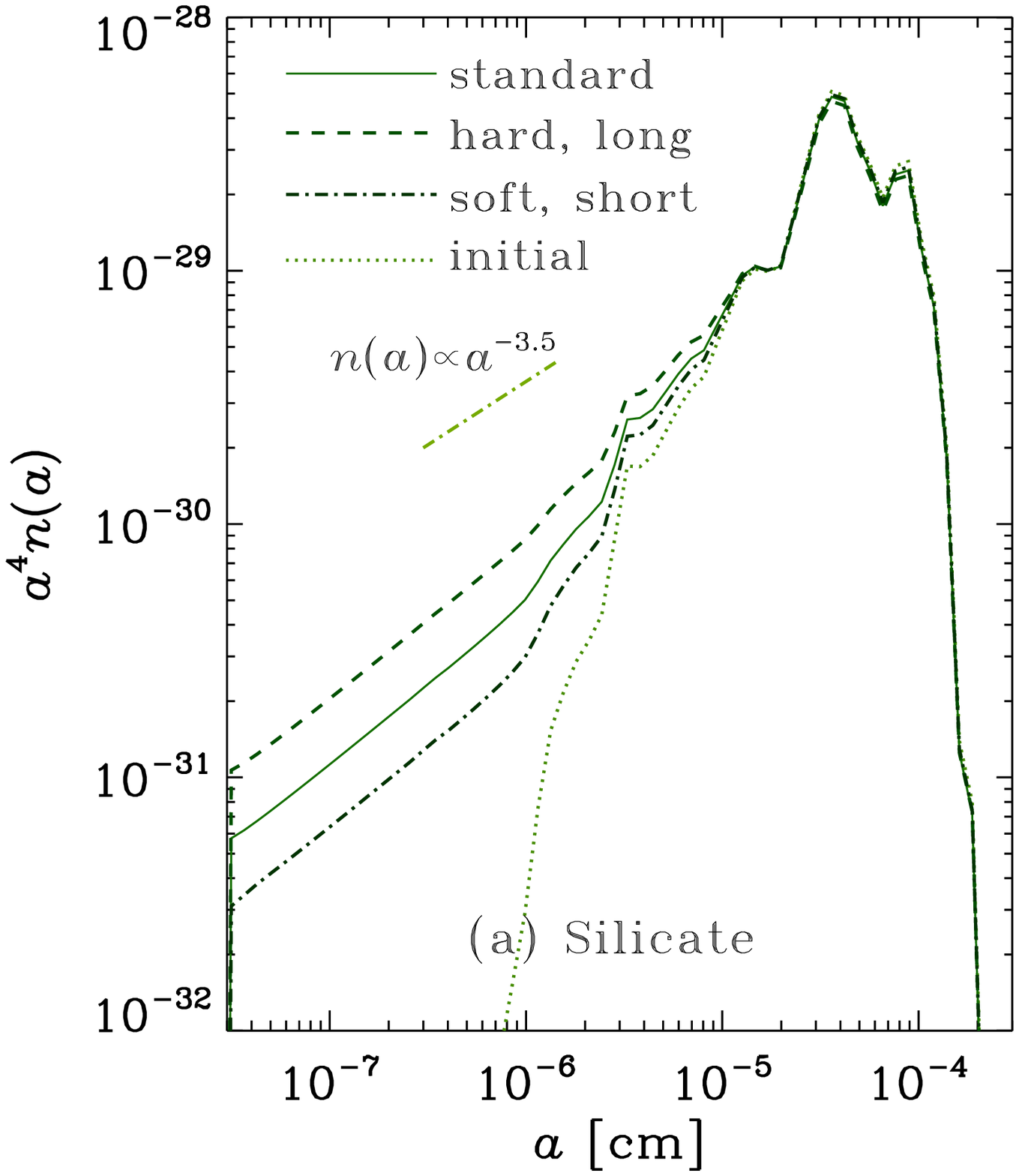}
\includegraphics[width=0.48\textwidth]{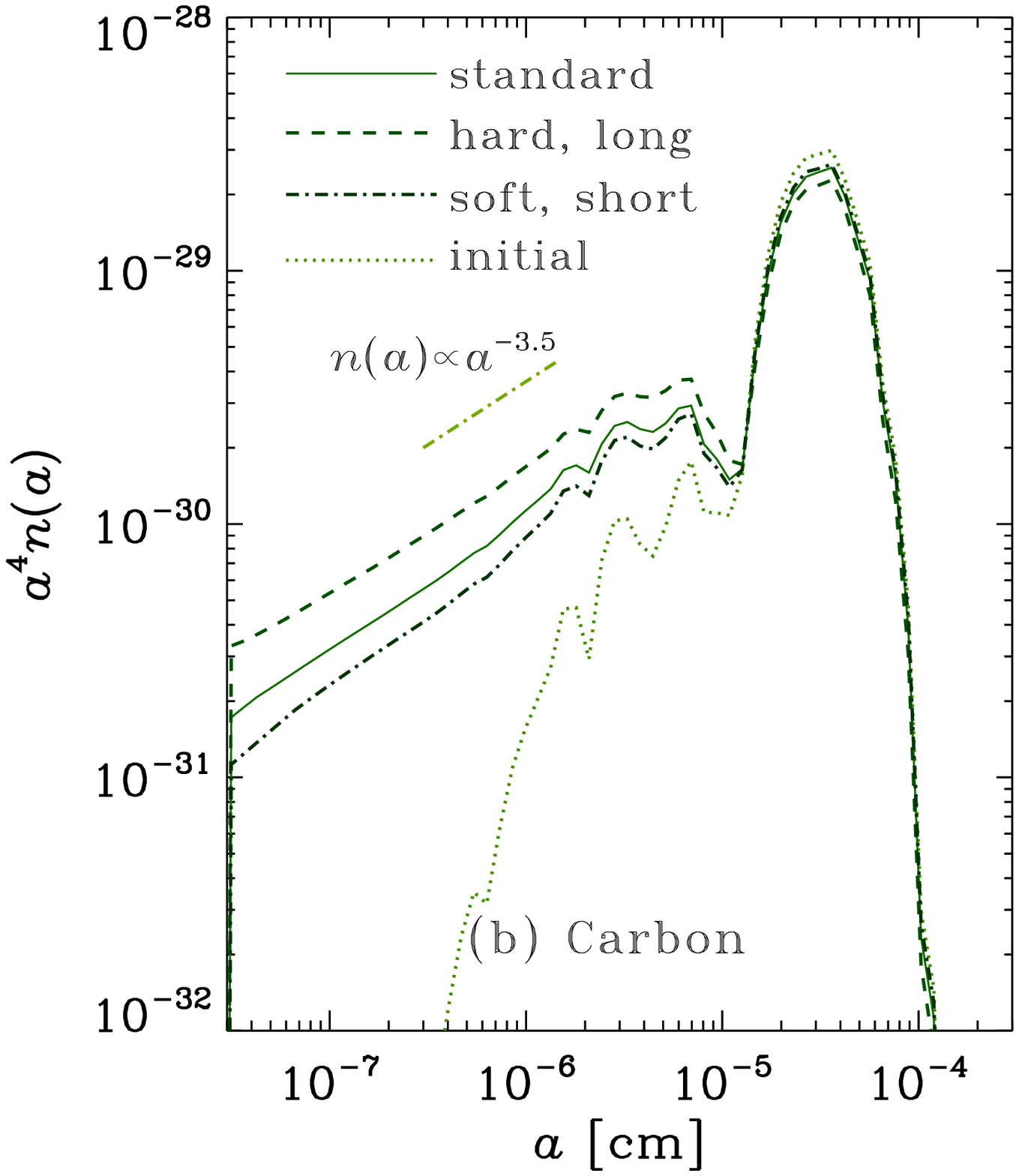}
\caption{\footnotesize
Same as Fig.\ \ref{fig:Pcr} but for constant $P_1/t$.
The solid, dashed, and dot-dashed lines show the results
for (a) 
$(P_1,\, t)=(3\times 10^{11}~\mathrm{dyn~cm}^{-1},\,
10~\mathrm{Myr})$, $(1\times 10^{11}~\mathrm{dyn~cm}^{-1},\,
3.3~\mathrm{Myr})$, and $(9\times 10^{11}~\mathrm{dyn~cm}^{-1},\,
30~\mathrm{Myr})$, respectively, for silicate, and (b)
$(P_1,\, t)=(4\times 10^{10}~\mathrm{dyn~cm}^{-1},\,
10~\mathrm{Myr})$, $(1.3\times 10^{10}~\mathrm{dyn~cm}^{-1},\,
3.3~\mathrm{Myr})$, and
$(1.2\times 10^{11}~\mathrm{dyn~cm}^{-1},\,
30~\mathrm{Myr})$, respectively, for carbonaceous dust.
The dotted line presents the initial
grain size distribution before shattering.
}
\label{fig:Pcr_test}
\end{figure*}

The resulting grain size distribution is sensitive
to various parameters in different ways.
As expected, shattering is more efficient for
a smaller value of $P_1$ simply because
the grains are softer. Since the cratering volume
is approximately proportional to $1/P_1$
(Jones \textit{et al}., 1996), it is expected
that similar grain size distributions are
realized if we keep $P_1/t$ constant.
To examine if this is the case, we compare the
resulting grain size distributions for constant
$P_1/t$ in Fig.~\ref{fig:Pcr_test}. We
examine
$(P_1,\, t)=(3\times 10^{11}~\mathrm{dyn~cm}^{-1},\,
10~\mathrm{Myr})$, $(1\times 10^{11}~\mathrm{dyn~cm}^{-1},\,
3.3~\mathrm{Myr})$ (i.e., soft dust and short shattering
duration), and $(9\times 10^{11}~\mathrm{dyn~cm}^{-1},\,
30~\mathrm{Myr})$ (i.e., hard dust and long shattering
duration) for silicate, and
$(P_1,\, t)=(4\times 10^{10}~\mathrm{dyn~cm}^{-1},\,
10~\mathrm{Myr})$, $(1.3\times 10^{10}~\mathrm{dyn~cm}^{-1},\,
3.3~\mathrm{Myr})$, and
$(1.2\times 10^{11}~\mathrm{dyn~cm}^{-1},\,
30~\mathrm{Myr})$ for carbonaceous dust.
We observe that similar grain size distributions are indeed
obtained within a factor of 3.
{The difference arises because the scaling
of the cratering volume (or $M_\mathrm{ej}$) with
$P_1^{-1}$ is not perfect
(Section \ref{subsec:shatter}).}

The power index of fragment size distribution,
$\alpha_\mathrm{f}$, is simply reflected in the resulting
grain size distribution for $\alpha_\mathrm{f}=4.3$.
However, if $\alpha_\mathrm{f}$ is smaller than 3.5
(e.g., the case with $\alpha_\mathrm{f}=2.3$),
the slope of grain size
distribution at $a\lesssim 10^{-5}$ cm continuously
``steepens'' and approaches to the slope consistent
with MRN [$n(a)\propto a^{-3.5}$]
(Fig.\ \ref{fig:alpha}).
According to the analytical
studies of shattering by Hellyer (1970),
the power index of grain size distribution after shattering
has a steady state value of $\simeq -3.5$ if the size
distribution of shattered fragments is assumed to be
a power law with an index smaller than 4
(i.e., $\alpha_\mathrm{f}<4$) (see also
Tanaka \textit{et al}., 1996), although we should
note that the relative velocity between grains is
a complicated function of grain sizes in our case.
For $\alpha_\mathrm{f}>4$, since the mass occupied by
the smallest fragments is dominating, the size
distribution after shattering is just dominated by
the smallest fragments, keeping a power law index
of $-\alpha_\mathrm{f}$ at small sizes.

In general, larger grains have larger velocity
dispersions if they are accelerated by turbulence,
primarily because they are coupled with
larger-scale turbulent motion which has larger
velocity dispersions. Thus, even if the initial
grain size distribution is biased to large grains
with $a\gtrsim 10^{-5}$ cm, these grains are
efficiently shattered to produce a large abundance
of small grains. If these shattered grains have also
larger velocities than shattering
threshold, they collide with each other,
accelerating the production of small grains. Thus,
as shown in Fig.\ \ref{fig:vel}, the effect of
$a_\mathrm{c}$ (or the critical grain radius
above which the grains have larger velocities than
the shattering threshold) can be seen in the grain size
distribution, causing a deviation from a
monotonic power-law-like functional form.
The imprint of $a_\mathrm{c}$ is more clear in
carbonaceous dust than in silicate dust, since the
former species is softer. Interestingly,
Weingartner and Draine (2001) have shown that
grain size distributions derived from the Milky
Way extinction curves are complicated for
carbonaceous dust, which may be
explained by imprints of $a_\mathrm{c}$.

\subsection{Implication for the evolution of grain
size distribution in galaxies}\label{subsec:implication}

Inoue (2011) shows theoretically that the overall
dust mass in metal-enriched systems such as
the Milky Way is regulated by the balance between
dust growth in molecular clouds and dust destruction
{by sputtering} in SN shocks.
While both these two mechanisms deplete small
grains (Hirashita and Nozawa, 2012), shattering is
a unique mechanism that reproduces the
small grains. Indeed, Hirashita and Nozawa (2012)
show that, unless
we consider shattering, we underproduce the small
grain abundance in the Milky Way (see also
Asano \textit{et al}., 2013). As shown in
Section \ref{subsec:frag}, shattering not only
produces small grains but also has an inherent
mechanism of steepening the grain size distribution;
that is, even if the grain size distribution of
shattered fragments has $\alpha_\mathrm{f}$
smaller than
3.5, the resulting grain size distribution
has a steeper dependence, approaching a
slope consistent with the one derived by
MRN [i.e., $n(a)\propto a^{-3.5}$]. Thus, the
robust prediction is that, if shattering is the
dominant mechanism of small grain production, the
grain size distribution approaches an MRN-like
grain size distribution as long as
$\alpha_\mathrm{f}\lesssim 3.5$. And if shattering
is the dominant mechanism of small grain production,
$\alpha_\mathrm{f}>4$ is rejected, because
the grain size distribution in the Milky
Way ISM has a power index of $-3.5$ (MRN).

In this paper, we have not included dust production
by AGB stars. Even if the dust supply from AGB stars
has a significant contribution, the results
in this paper are not altered as long as
the dust grains formed by AGB stars are biased
to large ($\gtrsim 0.1~\mu$m) sizes. Rather,
the importance of shattering is emphasized because
the additional dust production
by AGB stars enhances the dust abundance, making
grain--grain collision more frequent.
Production of
large grains from AGB stars is indicated {observationally
(Groenewegen, 1997;
Gauger \textit{et al.}, 1998;
Norris \textit{et al.}, 2012)
and theoretically
(H\"{o}fner, 2008; Mattsson \& H\"{o}fner, 2011)}.
We also neglected
grain growth in molecular clouds,
{which is suggested to be a major dust
formation mechanism even at high redshift
(Mattsson, 2011; Valiante \textit{et al.}, 2011).}
Grain growth, however, cannot be a
supplying mechanism of small grains.

For the shattering duration, since it is degenerate
with the critical pressure of grains (hardness of
grains), it is difficult to constrain it directly
from the models. If grains experience
shattering for several tens of Myr,
the grain size distribution possibly approaches
a power law whose power index is consistent with
the MRN, if $\alpha_\mathrm{f}\lesssim 3.5$
(Fig.\ \ref{fig:alpha_ev}).
Since the timescale of ISM phase exchange is
comparable or
shorter than that required for the grain size
distribution to
approach the MRN size distribution
(O'Donnell and Mathis, 1997), shattering may
occur intermittently. Even in this case, it is
expected that grains are relaxed into an MRN-like
simple power law if the total shattering duration
reaches several tens of Myr. Such a duration for
shattering is probable since it
is much shorter than the grain lifetime
($\sim\mbox{a few}\times 10^8$ Myr;
Jones \textit{et al}., 1996).

\section{Conclusion}\label{sec:conclusion}

Shattering is a viable mechanism of
small grain production in the ISM both
in nearby galaxies and in high-redshift galaxies.
We have examined if grain size distributions predicted
by shattering models
are robust against the change of various parameters
and formulations.
Because of the uncertainty in $f_M$ (the fraction
of the shocked material that is eventually ejected
as fragments), the predicted grain size distribution
after shattering is uncertain by a factor of 1.3
(1.6) for silicate (carbonaceous dust).
We have identified $P_1$
(the critical pressure above which {the original
lattice structure is destroyed})
as the most important quantity in determining the timescale
of small grain production, and confirmed that the same
$P_1/t$ ($t$ is the duration of shattering) gives roughly
the same
grain size distributions after shattering within a
factor of 3.

A simpler and more intuitive formulation
by KT10 is in good agreement with
our model based on Jones \textit{et al}.\ (1996) within
a factor of 1.8 (1.4) for silicate (carbonaceous dust)
if we focus on the small grain
production at $a\lesssim 10^{-6}$ cm. This is as small as
the uncertainty caused by $f_M$.
Thus, as long as we have an uncertainty in $f_M$,
it is sufficient to adopt the simpler formulation by KT10.
The size distribution of shattered fragments have minor
effects as long as $\alpha_\mathrm{f}\lesssim 3.5$, since
the grain size distribution is continuously steepened
by shattering and become consistent with the MRN grain
size distribution regardless of the value of
$\alpha_\mathrm{f}$ ($\lesssim 3.5$).

The effect of the grain velocities as a function of
grain radius can be seen {more clearly} in
carbonaceous grains than
in silicate, since the former species is shattered
more easily. Thus, it is predicted that carbonaceous
species has more complicated grain size distribution than
silicate; in other words, the grain size distribution
of carbonaceous dust can show some imprints of the
grain velocity as a function of grain radius.

\acknowledgments{
We are grateful to anonymous referees
for helpful comments. HH has been
supported through NSC grant 99-2112-M-001-006-MY3.
HK gratefully acknowledges the support from Grants-in-Aid from MEXT (23103005).
}


\email{H. Hirashita (e-mail: hirashita@asiaa.sinica.edu.tw)
and H. Kobayashi}

\label{finalpage}
\lastpagesettings

\end{document}